\documentclass[useAMS, usenatbib]{mnras}

\usepackage{times}
\usepackage{amssymb}
\usepackage{amsmath}
\usepackage{bm}
\usepackage{rotating}
\usepackage{epsfig}
\usepackage{graphicx}
\usepackage{grffile}

\input{epsf}

\title[Fe K$\alpha$ Compton shoulder]
{Sensitivity of the Fe K$\alpha$ Compton shoulder to the geometry and variability of the X-ray illumination of cosmic objects}

\author[H. Odaka et al.]
{Hirokazu Odaka$^{1,2}$, Hiroki Yoneda$^{3,2}$, Tadayuki Takahashi$^{2,3}$, Andrew Fabian$^4$\\
\\
$^1$ Kavli Institute for Particle Astrophysics and Cosmology, Stanford University, 2575 Sand Hill Rd, Menlo Park, CA, 94025, USA \\
$^2$ Institute of Space and Astronautical Science, JAXA, 3-1-1 Yoshinodai, Chuo-ku, Sagamihara, Kanagawa, Japan \\
$^3$ Department of Physics, University of Tokyo, 7-3-1 Hongo, Bunkyo-ku, Tokyo, Japan \\
$^4$ Institute of Astronomy, University of Cambridge, Madingley Road, Cambridge, UK \\
}

\date{Submitted to MNRAS}
\pagerange{\pageref{firstpage}--\pageref{lastpage}} \pubyear{}

\begin{document}

\topmargin = -0.5cm

\maketitle

\label{firstpage}

\begin{abstract}
In an X-ray reflection spectrum, a tail-like spectral feature generated via Compton down-scattering, known as a Compton shoulder (CS), appears at the low-energy side of the iron K$\alpha$ line.
Despite its great diagnostic potential, its use as a spectral probe of the reflector has been seriously limited due to observational difficulties and modelling complexities.
We revisit the basic nature of the CS by systematic investigation into its dependence on spatial and temporal parameters.
The calculations are performed by Monte-Carlo simulations for sphere and slab geometries.
The dependence is obtained in a two-dimensional space of column density and metal abundance, demonstrating that the CS solves parameter degeneration between them which was seen in conventional spectral analysis using photoelectric absorption and fluorescence lines.
Unlike the iron line, the CS does not suffer from any observational dependence on the spectral hardness.
The CS profile is highly dependent on the inclination angle of the slab geometry unless the slab is Compton-thick, and the time evolution of the CS is shown to be useful to constrain temporal information on the source if the intrinsic radiation is variable.
We also discuss how atomic binding of the scattering electrons in cold matter blurs the CS profile, finding that the effect is practically similar to thermal broadening in a plasma with a moderate temperature of $\sim$5 eV.
Spectral diagnostics using the CS is demonstrated with grating data of X-ray binary GX 301$-$2, and will be available in future with high-resolution spectra of active galactic nuclei obtained by microcalorimeters.
\end{abstract}

\begin{keywords}
X-ray reflection, Compton shoulder, Active galactic nuclei, black holes
\end{keywords}

\section{Introduction} \label{sec:introduction}


Accretion-powered objects such as black holes commonly display X-ray reflection from their environments in addition to intrinsic X-rays from the central engines.
For instance, reflection components from an accretion disc and a dusty torus around a supermassive black hole feature in the X-ray spectra of many active galactic nuclei (AGN).
In general, X-ray reflection is complex; with reprocessing by matter in the vicinity of the X-ray source via photoionisation and scattering.
The energy spectrum of X-ray reflection consists of fluorescence and recombination lines (when the reflector is a plasma) as well as a scattered and soft thermal continua \citep{Ross:2005, Garcia:2011}.
The reflection spectrum gives us important information both on the reflector itself and on the illuminating source.


Fluorescence lines (e.g., Fe K$\alpha$ at 6.4 keV) are remarkable features seen in the reflection spectrum.
These line photons can be Compton down-scattered to produce a low-energy tail associated with the main line \citep{George:1991, Matt:1991}.
This tail-like structure is called a Compton shoulder (CS).
The CS associated with Fe K$\alpha$ is the most prominent one because of the high abundance and high fluorescent yield of iron.
Since the CS is a result of Compton scattering of the monochromatic line photons, it should have a great diagnostic potential of the scattering medium.
However, to extract the physical properties of the scatterer is not a straightforward problem from both observational and theoretical aspects.


Observation of the CS requires high energy resolution owing to the small energy shift produced by the scattering, which is only 0.16 keV for Fe K$\alpha$ at 6.4 keV at maximum, i.e., for a scattering angle of $180^\circ$.
Only grating spectrometers have achieved sufficient resolution to measure the CS of Fe K$\alpha$ for point-like sources.
\textit{Chandra}-HETG, which has an energy resolution about 30 eV (full width at half maximum or FWHM), revealed a fully resolved CS from high-mass X-ray binary (HMXB) GX~301$-$2 which is generated in dense stellar wind \citep{Watanabe:2003}.
They obtained constraints on the wind properties including the column density, the metal abundance, and the electron temperature by comparing the data with detailed Monte-Carlo simulations.
Such a successful observation is still challenging due to limited energy resolution and/or inadequate collection area of conventional instruments.

An X-ray microcalorimeter is a promising detector technology that can change this situation.
\textit{Hitomi} also known as the X-ray Astronomy Satellite \textit{ASTRO-H} \citep{Takahashi:2014} achieved an unprecedented resolution of 4.9~eV (FWHM) at 6 keV and an improved collection area, which was an order of magnitude larger than that of \textit{Chandra}-HETG.
The microcalorimeter onboard \textit{Hitomi} clearly showed its expected performance by revealing the narrow line complex of He-like ion of iron around 6.6 keV from the core of the Perseus Cluster during the initial phase of the mission \citep{Hitomi:2016}.
Although \textit{Hitomi} was lost before planned observations of Compton-thick sources, the CS would undoubtedly become a useful spectral feature for characterising X-ray scattering medium in the microcalorimeter era \citep{Reynolds:2014, Smith:2014}.
Moreover, this spectral feature will be even more indispensable as a probe of Compton-thick AGN in late 2020's when \textit{Athena} \citep{Barcons:2015} brings us fine resolution data with high statistics from a much larger collection area.


The generation of the CS in dense matter is a complicated process from a viewpoint of spectral modelling.
\citet{Sunyaev:1996} reviewed the detailed physical processes and pointed out the importance of the state of electrons responsible for the scattering.
In an astrophysical plasma where most hydrogen and helium are ionised, most of the electrons exist as free electrons that have velocities sampled from the Maxwell--Boltzmann distribution specified by the electron temperature, and each scattering is governed by the Klein--Nishina differential cross section.
In cold matter, by contrast, the electrons are bound to atoms or molecules, leading to an altered scattering process, which causes a difference in the shape and intensity of the CS.
Atomic binding means that a finite momentum of the electron before scattering should be taken into account and that recoil of the electron can be suppressed.
It is therefore of great importance to consider the physical conditions of the electrons including whether they are free or bound to atoms or molecules.

A physically appropriate radiation model of X-ray reflection by cold matter has been developed in the context of X-ray reflection nebulae.
\citet{Sunyaev:1998} applied all the physical concepts of scattering to a problem of X-ray reflection of a past outburst of the central black hole in our galaxy by a molecular cloud, and discussed the time evolution of the spectrum and morphology of the reflected radiation including CSs.
More realistic cloud conditions were considered by \citet{Odaka:2011} using a Monte-Carlo simulation code called \texttt{MONACO}, which is able to treat accurate photon interactions in a complicated geometry, in order to interpret and predict X-ray reflection properties of the Sgr B2 cloud \citep{Sunyaev:1993, Koyama:1996, Revnivtsev:2004, Koyama:2008, Mori:2015, Zhang:2015} in the Galactic Centre region.
Since the CS should also be a promising probe for constraining properties of Compton-thick AGNs, the same simulation framework, \texttt{MONACO}, was applied to an AGN molecular torus that has clumpy structure to synthesise a spectral model of the torus reflection \citep{Furui:2016}.


The Monte-Carlo approach is generally suitable for accurate calculation of the reprocessed spectrum which can include a CS since it is capable of treating multiple interactions and competing processes including photoelectric absorption even in a complicated geometry.
Many authors have therefore adopted Monte-Carlo calculations to model X-ray reflection \citep[e.g.,][]{Leahy:1993, Murphy:2009, Ikeda:2009, Yaqoob:2012}.
Thus, in principle, it is now not difficult to produce a reflection spectrum including a CS by a detailed Monte-Carlo calculation once we fix all the physical conditions of the reflector and the illuminating source including their geometry.
Nonetheless, we again stress that detailed implementation of physical processes must be correct to obtain accurate results by Monte-Carlo simulations.


The CS, however, has not been established yet as a convenient observational probe owing to the complexity of the generation process.
It is not tractable to make full use of the CS for extracting information about the reflector and the X-ray source.
The spectral shape and the intensity of the CS depend on many factors such as the geometry, the optical depth and the chemical composition, preventing us from comprehensive investigation.
As a first step of deep understanding, it is highly beneficial to see quantitative behaviour of the CS in the case of simple geometry.
In this context, \citet{Matt:2002} studied sphere and slab geometries by Monte-Carlo simulations to obtain dependence on the column density, the metal abundance, and the viewing angle.


In the present paper, we revisit the basic nature of the CS as a measure of (at least modestly) Compton-thick object by means of Monte-Carlo simulations for simple geometries---a sphere and a slab---as adopted by \citet{Matt:2002} in order to understand its dependence upon properties of X-ray reflection system.
This theoretical calculation is of great importance for the accurate interpretation of high resolution data which is brought by grating spectrometers and microcalorimeters in the future.
In Section~\ref{sec:methodology}, we describe calculation methods and review basic physical concepts of the scattering processes.
We first consider one-dimensional spherical geometry in Section~\ref{sec:sphere}, investigating parameter dependence and correlation between parameters on the CS.
In Section~\ref{sec:slab}, we adopt a slab geometry since it is the possible simplest geometry to see angular dependence.
In these sections, we assume free electrons at rest as target electrons, which provide us with the simplest conditions, in order to depict the effects on the shape and intensity of the CS by the spatial and temporal properties of the reflector and the illuminating source.
Then, we demonstrate how additional complexities---finite temperature or atomic binding---change them in Section~\ref{sec:electron_states}.
We also discuss an application of the results to observational data in Section~\ref{sec:observation}. 
Section~\ref{sec:conclusions} summarises our conclusions.

\section{Methodology and Physical Processes}
\label{sec:methodology}

We adopt the Monte-Carlo approach to calculate precise spectra emerging from reflection.
This section describes basic methods of this study and physical processes related to the photon interactions used in the calculation.
In Section~\ref{subsec:code} we briefly write about the simulation code.
Section~\ref{subsec:physics} reviews photon processes which we should consider in the Monte-Carlo simulations.
In Section~\ref{subsec:analysis}, we describe a data analysis method by which spectral properties are extracted from a simulation spectrum.

\subsection{Simulation Framework}
\label{subsec:code}

For the Monte-Carlo simulations in the present work, we use the \texttt{MONACO} simulation code.
\texttt{MONACO} is designed as a general-purpose framework for calculating astrophysical X-ray radiation by treating radiative transfer based on the Monte-Carlo approach \citep{Odaka:2011}.
This framework utilises the \texttt{Geant4} toolkit library \citep{Agostinelli:2003, Allison:2006} for photon tracking in a complicated geometry.
Although \texttt{Geant4} contains implementation of physical processes related to the X-ray reflection, we do not use this built-in implementation but use our own implementation that is optimised for astrophysical applications so that we are able to treat the Doppler shifts and broadenings due to motion of reflecting matter.
We currently have three sets of physical processes in \texttt{MONACO}: (1) X-ray reflection by cold matter \citep{Odaka:2011}, (2) photon interactions in a photoionized plasma \citep{Watanabe:2006, Hagino:2015}, and (3) Comptonisation in a hot flow \citep{Odaka:2014}.
One of these physics sets should be selected according to an astrophysical situation of interest, and thus we use the first set in this work.

The framework is allowed to use any type of geometry that is supported by \texttt{Geant4}.
In this paper, we use a sphere in Section~\ref{sec:sphere} and a thin disc, which imitates a slab, in Section~\ref{sec:slab}.
X-ray reflection and absorption, which we should consider in this study, are physical processes without any dependence on an absolute spatial scale, and therefore an optical depth of the system is only essential.
We assume cosmic chemical composition by \citet{Anders:1989}, and we introduce a metal abundance $A_\mathrm{metal}$ relative to the cosmic value as a parameter that controls degree of photoelectric absorption due to metal elements (lithium and heavier elements).

\subsection{Physical Processes}
\label{subsec:physics}

\begin{table}
\caption{Cross sections at 6.4 keV of different scattering processes}
\begin{center}
\begin{tabular}{cc}
\hline\hline
Scattering process & Cross section [$\mathrm{cm^{-2}}$] \\
\hline
Rayleigh & $0.90\times 10^{-25}$ \\
Raman & $0.24\times 10^{-25}$ \\
Compton & $5.35\times 10^{-25}$ \\
Klein--Nishina (also called Compton) & $6.49\times 10^{-25}$ \\
Thomson (as reference) & $6.65\times 10^{-25}$ \\
\hline
\end{tabular}
\end{center}
\label{table:cross_section}
\end{table}%

\begin{figure}
\begin{center}
\includegraphics[width=8cm]{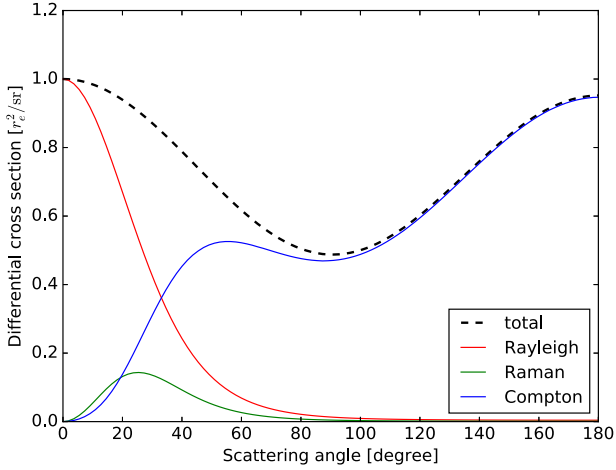}
\caption{Differential cross sections for the three scattering channels of a 6.4-keV photon by an electron bound to atomic hydrogen. The thin solid lines shows contributions of Rayleigh scattering (red), Raman scattering (green) and Compton scattering (blue). The thick dashed black line is sum of the total contributions. The values of the differential cross sections are normalized by the classical electron radius squared $r_\mathrm{e}{}^2$.}
\label{fig:dcs_hydrogen_atom}
\end{center}
\end{figure}

In cold matter in which all metals exist as neutral atoms, photon processes to be tracked are photoelectric absorption (and associated fluorescence) and scattering by an electron.
Section~\ref{subsec:physics_absorption} summarises implementation of the photoelectric absorption process.
The scattering process depends upon the electron state---whether the electrons are free or bound to atoms or molecules---determined by physical conditions such as temperature and radiation intensity.
Note that electrons responsible for scattering are mostly provided by hydrogen and helium, and contribution from heavier elements are negligible.
Physics of the scattering processes by a free electron and a bound electron is reviewed in \S\ref{subsec:physics_free_electron} and \S\ref{subsec:physics_bound_electron}, respectively.
Note that the same physics implementation were used for modelling of X-ray reflection nebulae in the Galactic Centre region \citep{Odaka:2011} and of AGN clumpy molecular tori \citep{Furui:2016}.

\subsubsection{Photoelectric Absorption}
\label{subsec:physics_absorption}

The physical process code for photoelectric absorption treats photoelectric absorptions and fluorescent line emissions following the absorptions.
After absorption by a K-shell electron, a K-shell fluorescent line photon is generated with the fluorescence yield, otherwise Auger electrons are emitted.
In our physics implementation, the fluorescence photon continues to be tracked, but the tracking calculation finishes for the Auger electron channel.
As cross section data of the photoelectric absorption, we adopt the Evaluated Photon Data Library 97 (EPDL97)\footnote{https://www-nds.iaea.org/epdl97/}, which is distributed together with the \texttt{Geant4} toolkit as data for electromagnetic processes at low-energy regime (below $\sim$1 MeV).
Though EPDL97 provides all necessary properties of the absorption process including fluorescence, we replace a part of the data regarding fluorescence with other databases that are more appropriate for X-ray astronomy as follows, since we find more accurate values.
Fluorescence yields, K-shell line energies, and K$\beta$-to-K$\alpha$ ratios are taken from \citet{Krause:1979}, \citet{Thompson:2001}, and \citet{Ertugral:2007}, respectively.

\subsubsection{Scattering by Free Electron}
\label{subsec:physics_free_electron}

The photon scattering process by a free electron is governed by Compton scattering.
Conservation of energy and momentum yields the relation between the photon energies before and after the scattering (the Compton scattering formula)
\begin{equation}\label{eq:compton}
h\nu_1 = \dfrac{h\nu_0}{1+\dfrac{h\nu_0}{m_ec^2}(1-\cos\theta)},
\end{equation}
where $h\nu_0$ and $h\nu_1$ are photon energies before and after the scattering ($\nu$ denotes a frequency and $h$ is the Planck constant), $m_e$ is the mass of an electron, $c$ is the speed of light, and $\theta$ is a scattering angle.
Subscripts 0 and 1 are used to indicate the states before and after the scattering, respectively.
The differential cross section of Compton scattering is given by Klein--Nishina's formula
\begin{equation}\label{eq:klein-nishina}
\frac{d\sigma}{d\Omega} = \frac{r_e{}^2}{2}\left(\frac{h\nu_1}{h\nu_0}\right)^2 \left(\frac{h\nu_0}{h\nu_1}+\frac{h\nu_1}{h\nu_0}-\sin^2\theta\right),
\end{equation}
where $r_e$ is the classical electron radius.
Equations (\ref{eq:compton}) and (\ref{eq:klein-nishina}) completely determine the spectral profile of a CS if target electrons of scattering are at rest.
If an initial photon energy is sufficiently low compared with the electron mass energy, $h\nu_0 \ll m_ec^2$, the differential cross section is reduced to
\begin{equation}\label{eq:thomson}
\left(\frac{d\sigma}{d\Omega}\right)_\mathrm{T} = \frac{r_e{}^2}{2}\left(\frac{h\nu_1}{h\nu_0}\right)^2 \left(1+\cos^2\theta\right),
\end{equation}
which is known as Thomson scattering.

In most astrophysical environments, target electrons are not necessarily at rest, and the motion of a target electron changes the CS spectral profile through the Doppler effect.
In simulations by \texttt{MONACO}, we treat effects of bulk motion and random motion by an algorithm using the Lorentz transformation \citep[see Appendix of][]{Odaka:2014}.
There may be thermal motion in addition to dynamical motion, which is also treated as random motion that is described by the Maxwell--Boltzmann velocity distribution.
In this paper, we do not treat bulk motion and do consider only thermal motion at moderate temperature of $\sim$5 eV, which means that relativistic effects are not significant.

\subsubsection{Scattering by Bound Electron}
\label{subsec:physics_bound_electron}

If a target electron is bound to an atom or a molecule, the scattering process is altered by the effect of binding.
A detailed review about this process is found in \citet{Sunyaev:1996}, and we here describe key concepts and physics that are implemented in the calculation code.
The scattering process by a bound electron can be classified into three channels according to the destination of the target electron; namely, the final state of the electron can be the ground state, excited states, or free states.
The channels are:
\begin{equation}\label{eq:scattering_bound}
\gamma_0+\mathrm{X}_{i} \to \left\{ \begin{array}{ll}
\gamma_1+\mathrm{X}_{i} & \text{(Rayleigh scattering)} \\
\gamma_1+\mathrm{X}^*_{i} & \text{(Raman scattering)} \\
\gamma_1+\mathrm{X}_{i-1} + \mathrm{e}^{-} & \text{(Compton scattering)}
\end{array}\right.
\end{equation}
where $\gamma_0$ and $\gamma_1$ denote photons before and after the scattering, respectively, $\mathrm{X}_i$ is an atom with the number $i$ of electrons, and an asterisk denotes an excited state.

We now consider a photon scattered by an electron bound to a hydrogen atom.
Through this process, the quantum state of the electron changes from an initial state $|i\rangle$ to a final state $|f\rangle$.
The doubly differential cross section is given by
\begin{equation}
\begin{split}
\frac{d^2\sigma}{d\Omega dh\nu_1}=&
r_e{}^2
 \left(\frac{h\nu_1}{h\nu_0}\right) (\bm{e}_0\cdot\bm{e}_1)^2  \\
&\times \sum_f \left| \langle f|e^{-i\bm{\chi}\bm{r}}|i\rangle \right|^2
\delta(\Delta E+\Delta h\nu),
\end{split}
\end{equation}
where
\begin{gather*}
\Delta E=E_f - E_i,\quad \Delta h\nu = h\nu_1 - h\nu_0, \\
\quad \bm{\chi}=\bm{q}/\hbar, \quad \bm{q}=\bm{k}_1 - \bm{k}_0,
\end{gather*}
$E_i$ and $E_f$ are the initial and the final energies of the electron;
$\bm{e}_0$ and $\bm{e}_1$ are unit vectors of polarisation.
The momentum transfer through this scattering is denoted by $\bm{q}$; $\bm{k}_0$ and $\bm{k}_1$ are the initial and the final momenta of the photon.

We have analytical solutions of the electron wave function for atomic hydrogen.
Thus, the differential cross sections of the three channels are given by analytical formulae.
For Rayleigh scattering, the differential cross section is written as
\begin{equation}\label{eq:diff_cs_rayleigh}
\frac{d\sigma}{d\Omega}=
\left( \frac{d\sigma}{d\Omega} \right)_\mathrm{T}
\left[ 1+ \left( \frac{1}{2}qa \right)^2 \right]^{-4},
\end{equation}
where $(d\sigma/d\Omega)_\mathrm{T}$ is given by Equation~(\ref{eq:thomson}), and
$a$ is $r_\mathrm{B}/\hbar$, $r_\mathrm{B}=\hbar/m_e c\alpha$ is the Bohr radius ($\hbar=h/2\pi$, $\alpha$ is the fine structure constant).
Since the electron state does not change through the Rayleigh scattering, the photon energy also remains unchanged, i.e., $h\nu_1 = h\nu_0$.

If a scattering results in excitation of an electron level that has a principal quantum number $n$, this process is called Raman scattering, and the differential cross section is given by
\begin{equation}
\begin{split}
\frac{d\sigma}{d\Omega}=&
\left( \frac{d\sigma}{d\Omega} \right)_\mathrm{T}
\frac{2^8}{3}\frac{(qa)^2}{n^3}
\left[ 3(qa)^2+ \frac{n^2-1}{n^2}\right] \\
&\times
\frac{\left[ (n-1)^2 / n^2 + (qa)^2 \right]^{n-3} }{\left[ (n+1)^2/n^2 + (qa)^2 \right]^{n+3} }.
\end{split}
\end{equation}
The energy shift of the scattered photon is equal to the transition energy of the electron, i.e., $\Delta h\nu = -\Delta E = -(E_f-E_i)$.
Note that substituting $n=1$ to this cross section expression reduces it into Equation~(\ref{eq:diff_cs_rayleigh}).

If the target electron is ionised through the scattering, we usually call the process Compton scattering.
This is the same name for the scattering by a free electron discussed in \S\ref{subsec:physics_free_electron} since a scattering process that causes an electron recoil is called Compton scattering, regardless of whether the electron is bound or free.
However, we should be careful about the difference between the two processes due to the electron binding effect.
In the whole cross section that the free electron would have, a certain fraction goes to Rayleigh scattering and Raman scattering due to insufficient recoil energy to ionise the electron, and the rest corresponds to Compton scattering, as shown in Table~\ref{table:cross_section}.
The energy of the photon after Compton scattering is not determined uniquely by the scattering angle since the initial electron has a finite momentum in the atomic system which broadens the scattered photon energy.
Consequently, the doubly differential cross section is written as
\begin{equation}
\frac{d^2\sigma}{d\Omega dh\nu_1} = 
\left( \frac{d\sigma}{d\Omega} \right)_\mathrm{T}
\left(\frac{h\nu_0}{h\nu_1}\right)
H_{fi}{}^2,
\end{equation}
where
\begin{equation}
\begin{split}
H_{fi}{}^2 =& 2^8 a^2 m(1-e^{-2\pi /pa})^{-1} \\
&\exp\left[ \frac{-2}{pa}\tan^{-1}\left(\frac{2pa}{1+q^2 a^2 -p^2 a^2}\right) \right] \\
&\times \left[ q^4 a^4 + \frac{1}{3}q^2 a^2 (1+p^2 a^2) \right] \\
& \times [(q^2 a^2 + 1 - p^2 a^2)^2 + 4p^2 a^2]^{-3}, \\
p^2/2m =& -\Delta h\nu-E_b.
\end{split}
\end{equation}
Here, $p$ and $E_b$ are the momentum of the ejected electron and the binding energy of the electron, respectively.
The energy difference of the photon, $|\Delta h\nu|$, should be larger than the binding energy $E_\mathrm{b}=13.6$ eV for Compton scattering.

Figure~\ref{fig:dcs_hydrogen_atom} shows the differential cross sections for the three scattering channels for a photon with the iron K$\alpha$ line energy, 6.4 keV.
At small scattering angles, an energy transfer to the target electron is not sufficient to ionise it, so that Rayleigh scattering and Raman scattering are dominant.
Compton scattering becomes dominant as the scattering angle increases. 
It is worth noting that the sum of the three channels is equal to the differential cross section of scattering by a free electron, which is given by the Klein--Nishima formula (\ref{eq:klein-nishina}).

Hydrogen may exist as a molecule $\mathrm{H}_2$ in a dense cloud.
If the target electron is bound to an $\mathrm{H}_2$ molecule, the scattering process is slightly different from that of an atomic hydrogen $\mathrm{H}_1$.
\citet{Sunyaev:1999} discusses treatment of the process for molecular hydrogen in astrophysical situations, and we adopt their treatment in \texttt{MONACO}.
The most important difference from that of atomic hydrogen is coherence by the two electrons which doubles the cross section per electron of Rayleigh scattering.
For inelastic scattering (Raman scattering and Compton scattering), we use the same physics implementation for atomic hydrogen described above as a good approximation.
We can completely neglect rotational and vibrational levels of the molecule in the X-ray band.
Calculations assuming the molecular form are found in our previous work of X-ray reflecting molecular clouds \citep{Odaka:2011} and of AGN tori \citep{Furui:2016}, though its effect on the CS properties is negligible.
Since this paper focuses on the basic nature of CSs, we do not consider molecular hydrogen hereafter.

Scattering by an electron bound to helium should also be considered in most astrophysical situations.
Since an atomic helium has two electrons like a molecular hydrogen, Rayleigh scattering per electron is enhanced by a factor of 2 due to the coherent effect.
We use numerical calculations of differential cross sections provided by \citet{Vainshtein:1998}.
Helium has a larger binding energy $E_\mathrm{b}=24.6$ eV and larger momentum of bound electrons than those of hydrogen, which affects a spectral profile of scattered photons.

\subsection{Extraction of Spectral Quantities}
\label{subsec:analysis}

\begin{figure}
\begin{center}
\includegraphics[width=9.0cm]{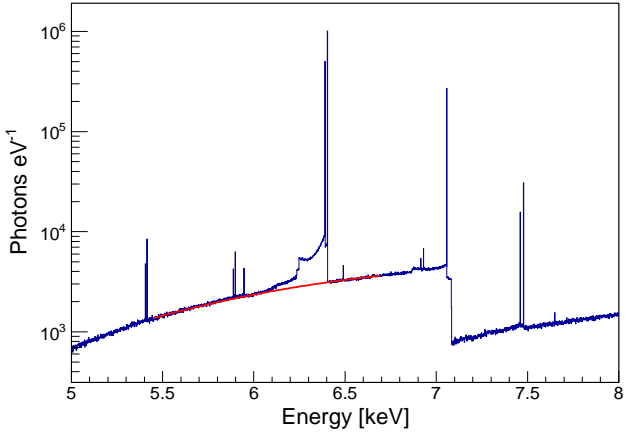}
\caption{A spectrum around the iron lines (K$\alpha_1$, K$\alpha_2$) at 6.4 keV and their CS emerging from a spherical cloud with $N_\mathrm{H} = 10^{24}\;\mathrm{cm^{-2}}$ and $A_\mathrm{metal} = 1.0$. A function fitted to the continuum is shown as a red line.}
\label{fig:example-spectrum}
\end{center}
\end{figure}

Figure~\ref{fig:example-spectrum} shows a spectrum around the iron line and its CS emerging from a spherical cloud with a column density of $N_\mathrm{H} = 10^{24}\;\mathrm{cm^{-2}}$ and a metal abundance of $A_\mathrm{metal} = 1.0$.
A simulation to obtain this spectrum is described in Section~\ref{sec:sphere}.
We here explain how to extract spectral quantities that characterise the CS by using this spectrum as an example.
To evaluate the continuum level, we fit a curved power law function
\begin{equation}
\frac{dN(E)}{dE} = p_0 x^{p_1+p_2 x}, \quad x = \frac{E}{6.4\ \mathrm{keV}},
\end{equation}
to the spectrum in energy ranges where contributions from the lines and the shoulder are negligible.
In the fitting, $p_0$, $p_1$, and $p_2$ are treated as free parameters.
The energy ranges for the fitting to the continuum are $5.44\;\mathrm{keV}<E<5.69\;\mathrm{keV}$ and $6.42\;\mathrm{keV}<E<6.70\;\mathrm{keV}$.

Several quantities obtained from the spectrum are useful for characterising the reflection spectrum which has a CS.
The equivalent widths (EWs) of the iron line ($\mathrm{K\alpha_1}$ and $\mathrm{K\alpha_2}$) and the CS are defined as
\begin{gather}
\label{eq:definition_ew_ka}
\mathrm{EW_{K\alpha}} \equiv \int_{E=6.086\;\mathrm{keV}}^{E=6.404\;\mathrm{keV}}\frac{S(E)-C(E)}{C(E)}dE, \\
\label{eq:definition_ew_cs}
\mathrm{EW_{CS}} \equiv \int_{E=6.086\;\mathrm{keV}}^{E=6.390\;\mathrm{keV}} \frac{S(E)-C(E)}{C(E)}dE,
\end{gather}
where $S(E)$ and $C(E)$ are the spectrum data as a function of energy and the fitted function to the continuum, respectively.
The upper limits of these integrals are set immediately above Fe K$\alpha_1$ (6.40384 keV) for $\mathrm{EW_{K\alpha}}$ and immediately below Fe K$\alpha_2$ (6.39084 keV) for $\mathrm{EW_{CS}}$.
Note that $\mathrm{EW_{K\alpha}}$ includes contribution from the CS above 6.086 keV, which is the minimum energy value of the second-order (scattered twice) CS of Fe K$\alpha_2$.
A fraction of the CS in the whole line is also a useful measure, which is defined as
\begin{equation}
f_\mathrm{CS} \equiv \dfrac{\displaystyle\int_{E=6.086\;\mathrm{keV}}^{E=6.390\;\mathrm{keV}} (S(E)-C(E))dE}{\displaystyle\int_{E=6.086\;\mathrm{keV}}^{E=6.404\;\mathrm{keV}}(S(E)-C(E))dE}.
\end{equation}
In addition, we introduce a measure $\Lambda$ to characterise a shape of a shoulder as
\begin{gather}
\Lambda \equiv \left(\int_{E=\mathrm{6.247\;keV}}^{E=\mathrm{6.390\;keV}}\left(E-E_c\right)^2 s(E)dE\right)^{\frac{1}{2}}, \\
E_c \equiv \frac{\mathrm{6.247\;keV+6.390\;keV}}{2}=\mathrm{6.3185\;keV}, \\
s(E)=\dfrac{S(E)-C(E)}{\displaystyle\int_{E=6.247\;\mathrm{keV}}^{E=6.390\;\mathrm{keV}}(S(E)-C(E))dE}.
\end{gather}
It is possible to regard $\Lambda$ as a statistical deviation that uses the midpoint of the energy band (6.247--6.390 keV) as the central value, but it is not a standard deviation.
The upper limit of this energy range remains the same as Equation (\ref{eq:definition_ew_cs}), but the lower limit is newly defined as the minimum energy value of the first-order (scattered once) CS of Fe K$\alpha_1$ since we want to characterise the main component of the CS profile.

\section{Spherical Geometry}
\label{sec:sphere}

\begin{figure}
\begin{center}
\includegraphics[width=5.0cm]{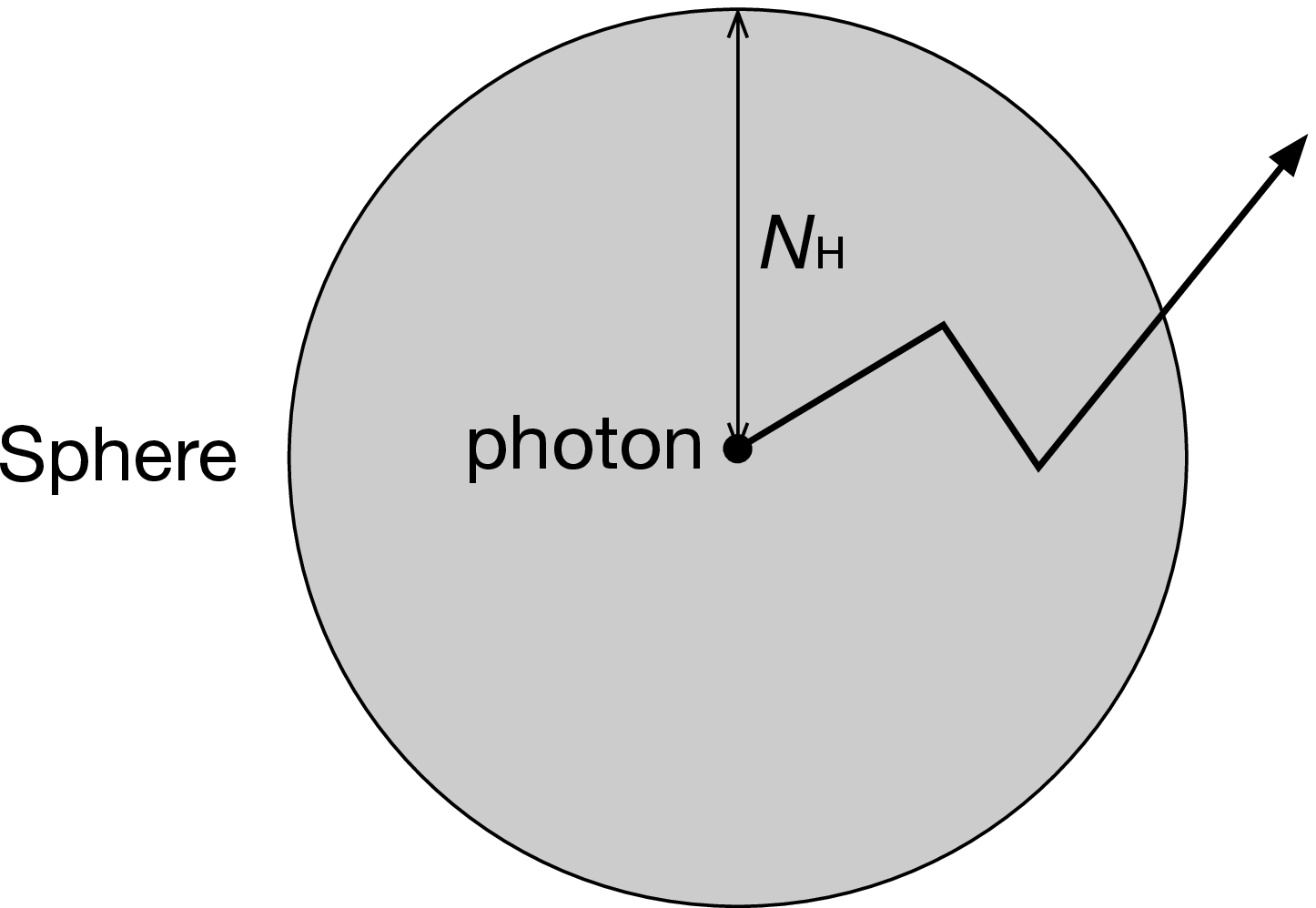}\\
\vspace{0.5cm}
\includegraphics[width=6.4cm]{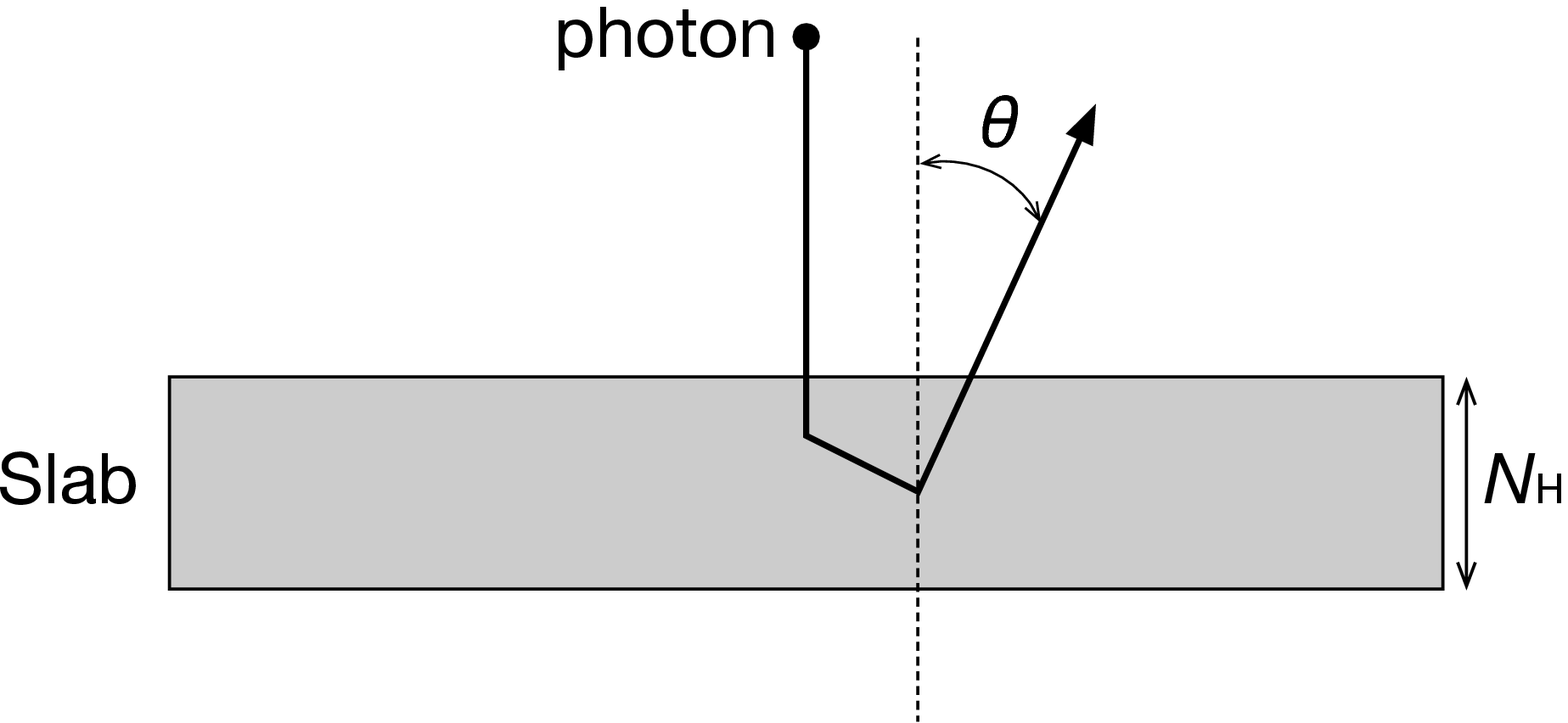}
\caption{Cross section view of geometries which we use in this work. Top: a sphere geometry. An X-ray emitter is positioned at the centre of the sphere, and an X-ray direction is sampled isotropically from a solid angle of 4$\pi$. Bottom: a disc geometry emulating a slab. An X-ray emitter is put above the disc and the X-ray direction is fixed to be vertically downward, i.e., $(0,\;0,\;-1)$. The disc radius is sufficiently large to extinguish effects of the edge. The viewing angle $\theta$ is defined by the angle between the disc normal and the photon escaping direction.}
\label{fig:geometry}
\end{center}
\end{figure}

We start with a spherical cloud to see the basic nature of the CS, as shown in the upper panel of Figure~\ref{fig:geometry}.
An X-ray source is positioned at the centre of the spherical cloud.
Since this geometry is essentially one-dimensional, it is suitable for studying dependence on basic parameters such as a column density, a metal abundance, and a spectral slope of an illuminating source.
The cloud has only two parameters, a hydrogen column density $N_\mathrm{H}$ measured from the centre to the surface, and a metal abundance $A_\mathrm{metal}$.
For each parameter set $(N_\mathrm{H},\;A_\mathrm{metal})$, we performed a simulation that generated $2\times 10^{8}$ primary photons sampled from a power law spectrum with a photon index of 2.0 in an energy range of 5--100 keV.
For several parameter sets that result in very thin optical depths or heavy absorption, we needed to run simulations with $2\times 10^{9}$ primary photons to obtain enough statistics.

\subsection{Dependence on column density and metal abundance}
\label{subsec:sphere_dependence_NH_MA}

\begin{figure*}
\begin{center}
\includegraphics[width=7.5cm]{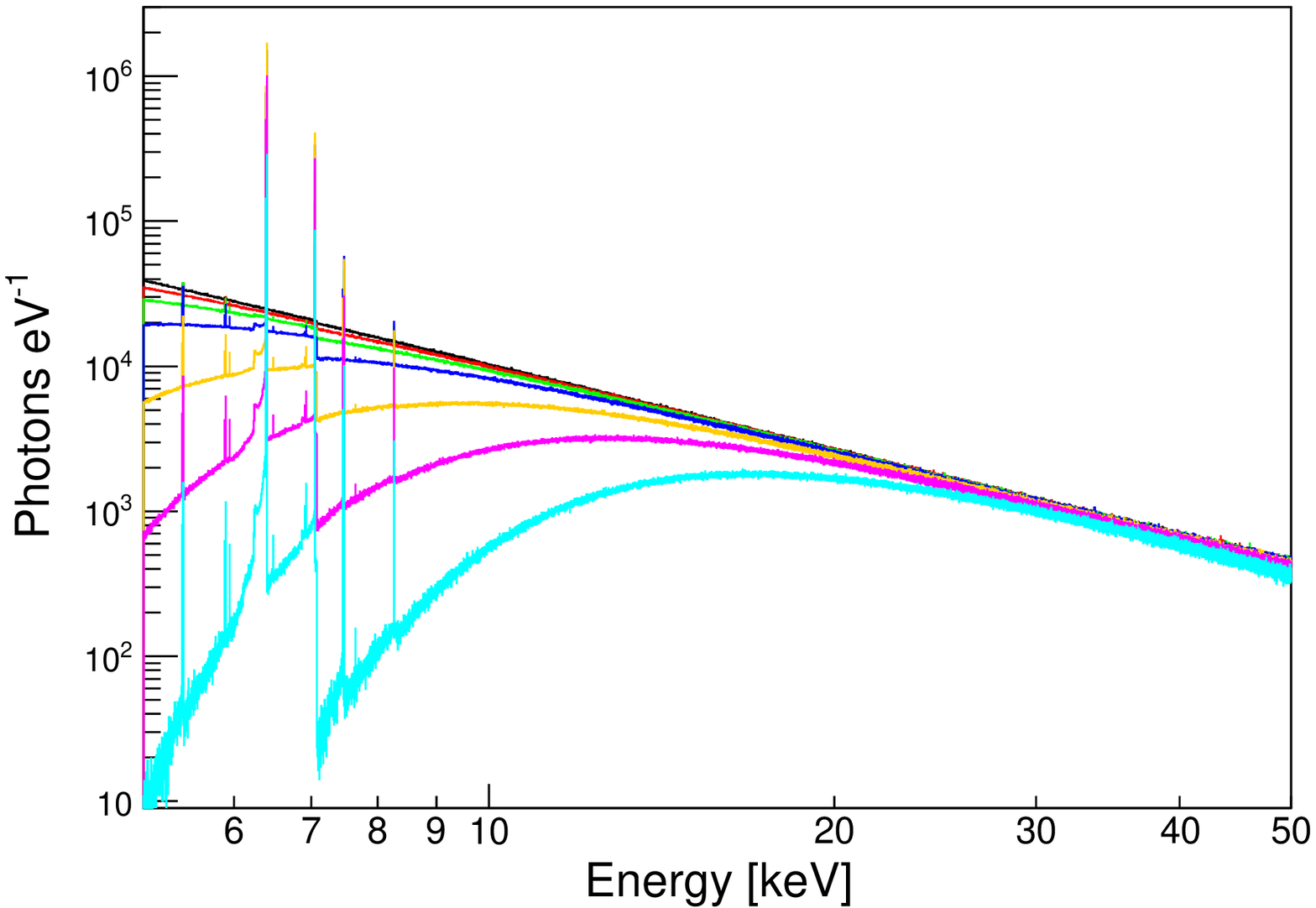}
\includegraphics[width=7.5cm]{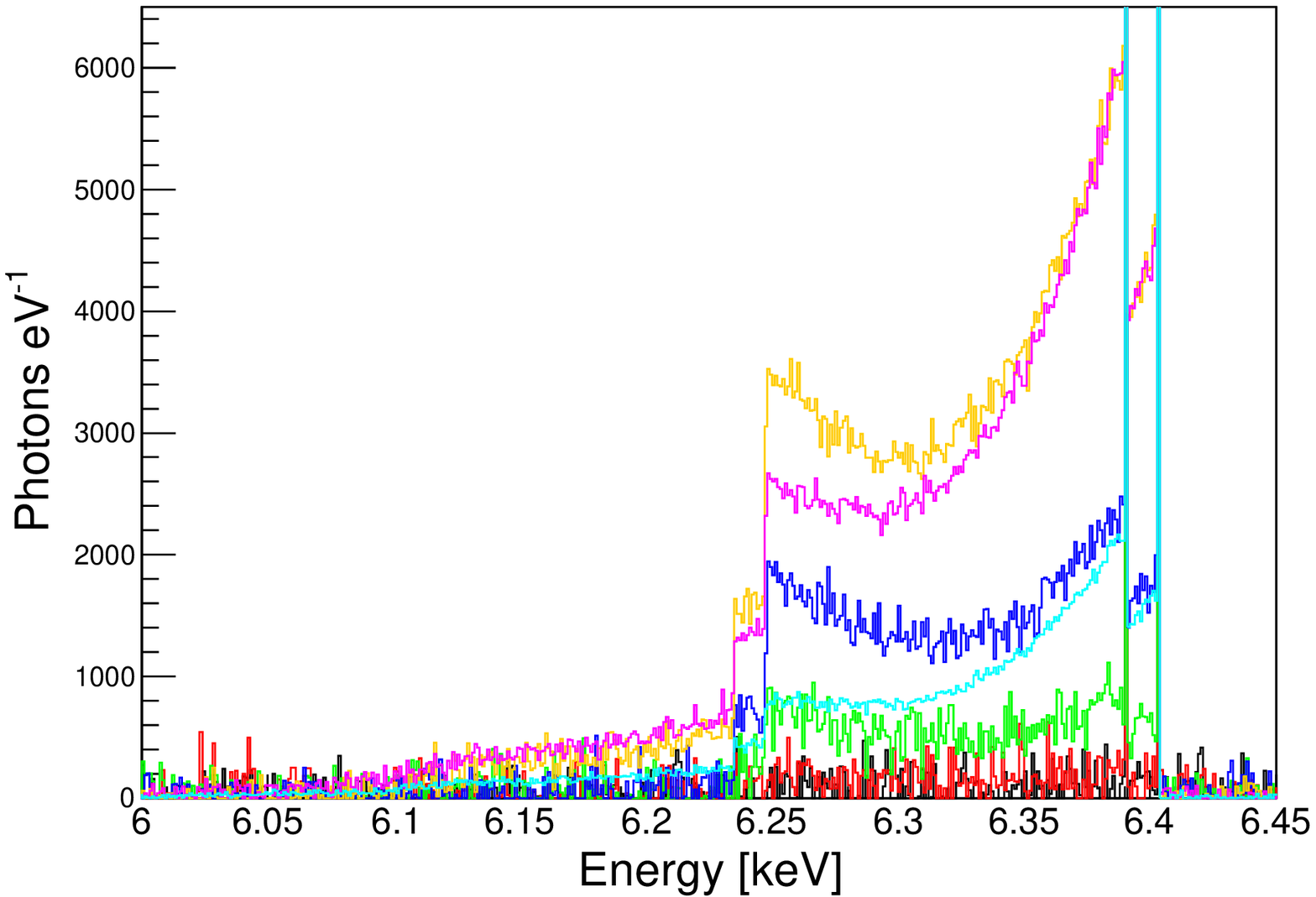} \\
\includegraphics[width=7.5cm]{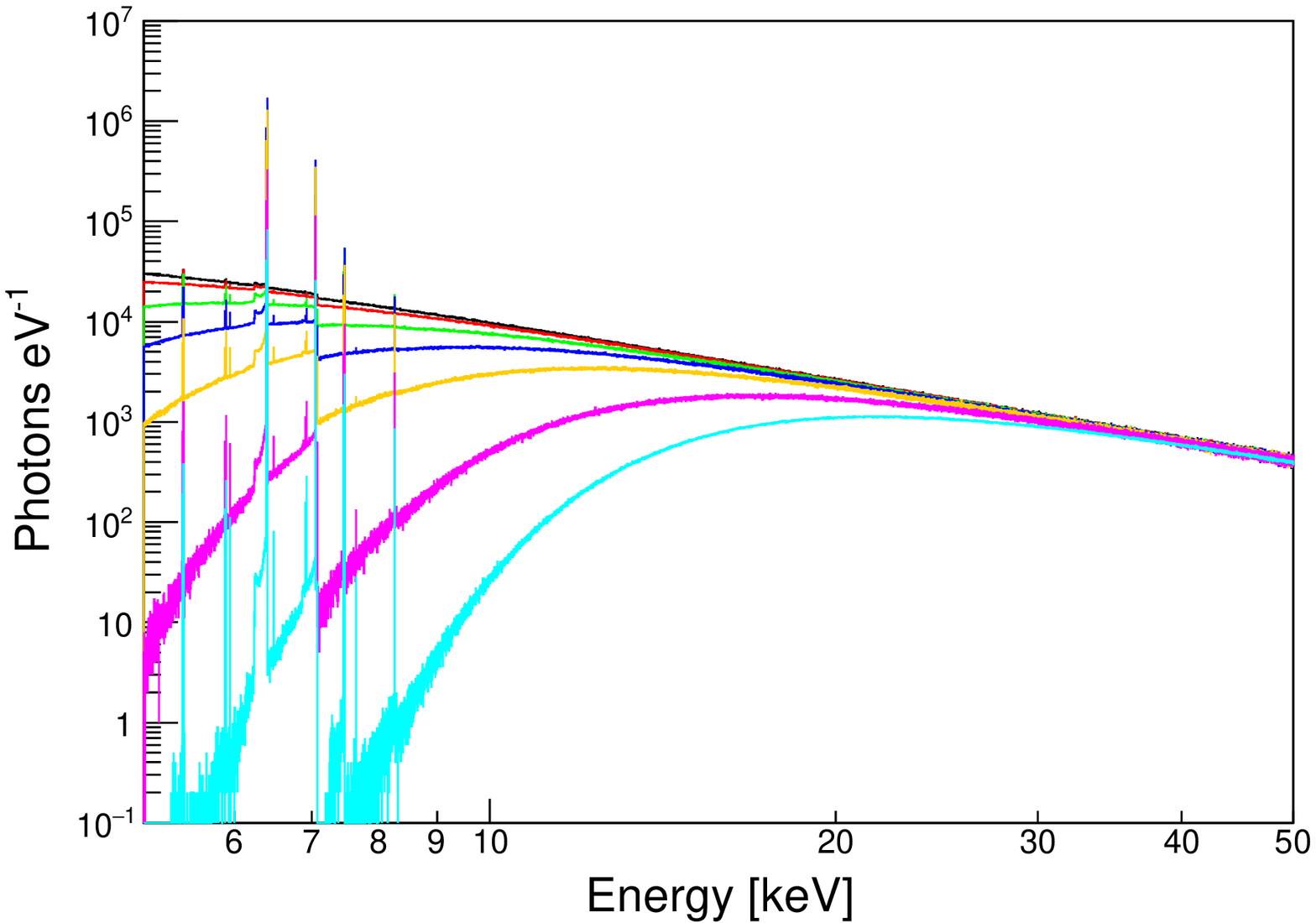}
\includegraphics[width=7.5cm]{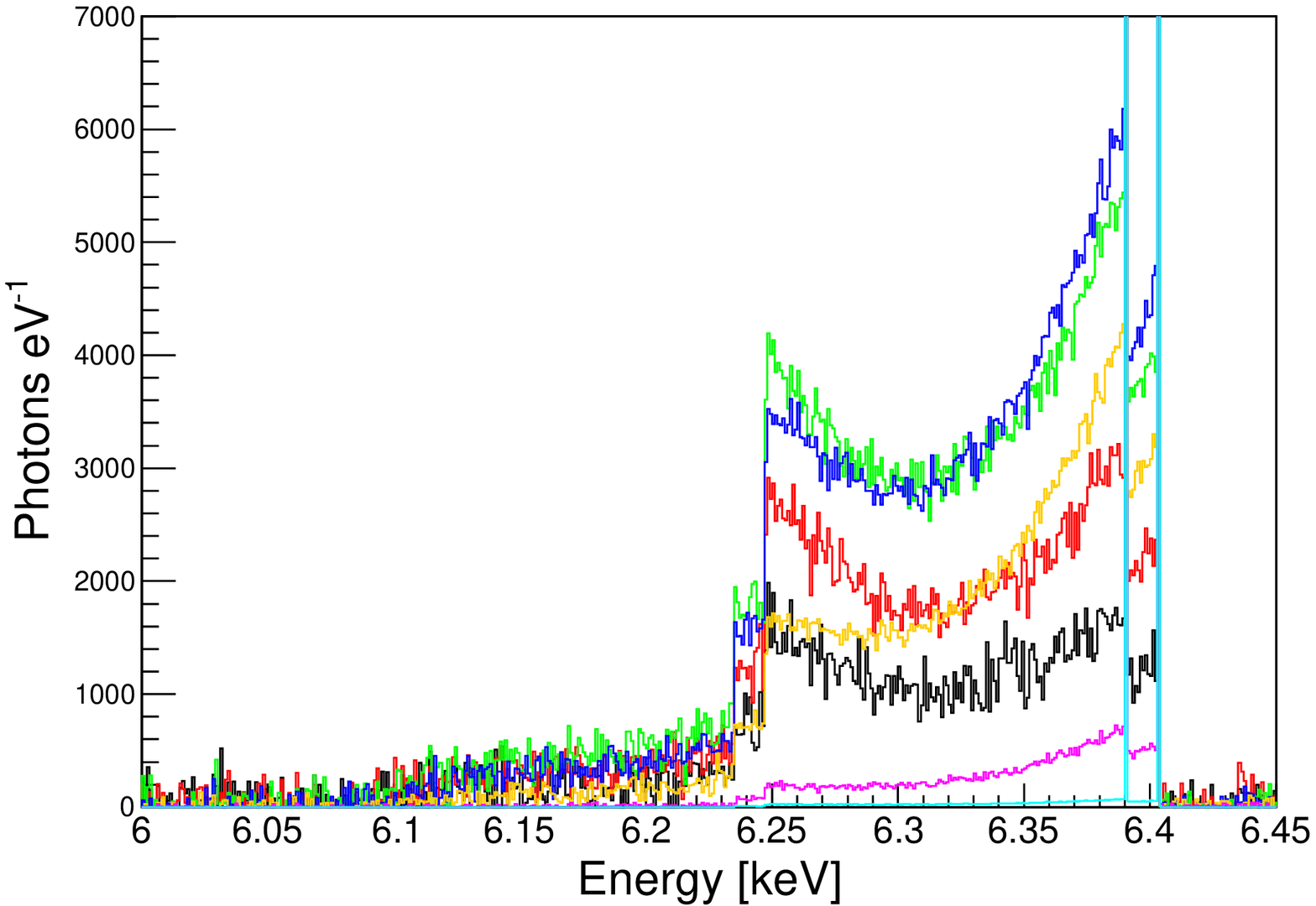}
\caption{X-ray spectra emerging from spherical clouds with different cloud parameters. Broadband spectra are shown in the left column, and spectra enlarged around the iron line in the right column where the underlaying continua are subtracted for clear comparison of the CS profile.
In the upper panels, we apply different column densities of $N_\mathrm{H}=2\times 10^{22}\;\mathrm{cm^{-2}}$ (black), $5\times 10^{22}\;\mathrm{cm^{-2}}$ (red), $1\times 10^{23}\;\mathrm{cm^{-2}}$ (green), $2\times 10^{23}\;\mathrm{cm^{-2}}$ (blue), $5\times 10^{23}\;\mathrm{cm^{-2}}$ (yellow), $1\times 10^{24}\;\mathrm{cm^{-2}}$ (magenta), and $2\times 10^{24}\;\mathrm{cm^{-2}}$ (cyan), but a common value of the metal abundance $A_\mathrm{metal}=1.0$ is assumed.
In the bottom panels, we fix $N_\mathrm{H}=5\times 10^{23}\;\mathrm{cm^{-2}}$ and change the metal abundance to $A_\mathrm{metal}=0.1$ (black), $0.2$ (red), $0.5$ (green), $1.0$ (blue), $2.0$(yellow), $5.0$ (magenta), and $10.0$ (cyan).
}
\label{fig:sphere_NH_MA_spectra}
\end{center}
\end{figure*}

\begin{figure*}
\begin{center}
\includegraphics[width=7.75cm]{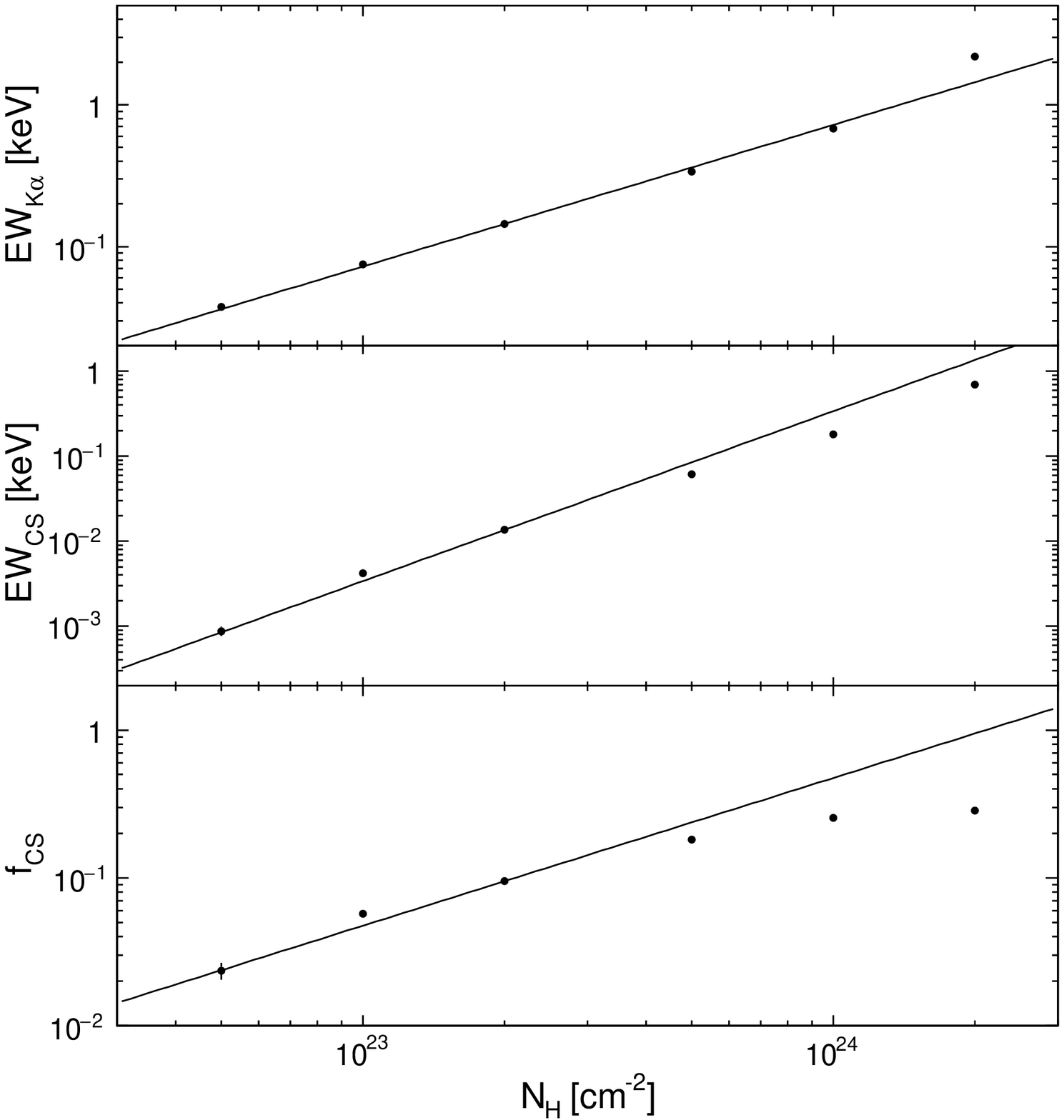}
\includegraphics[width=7.75cm]{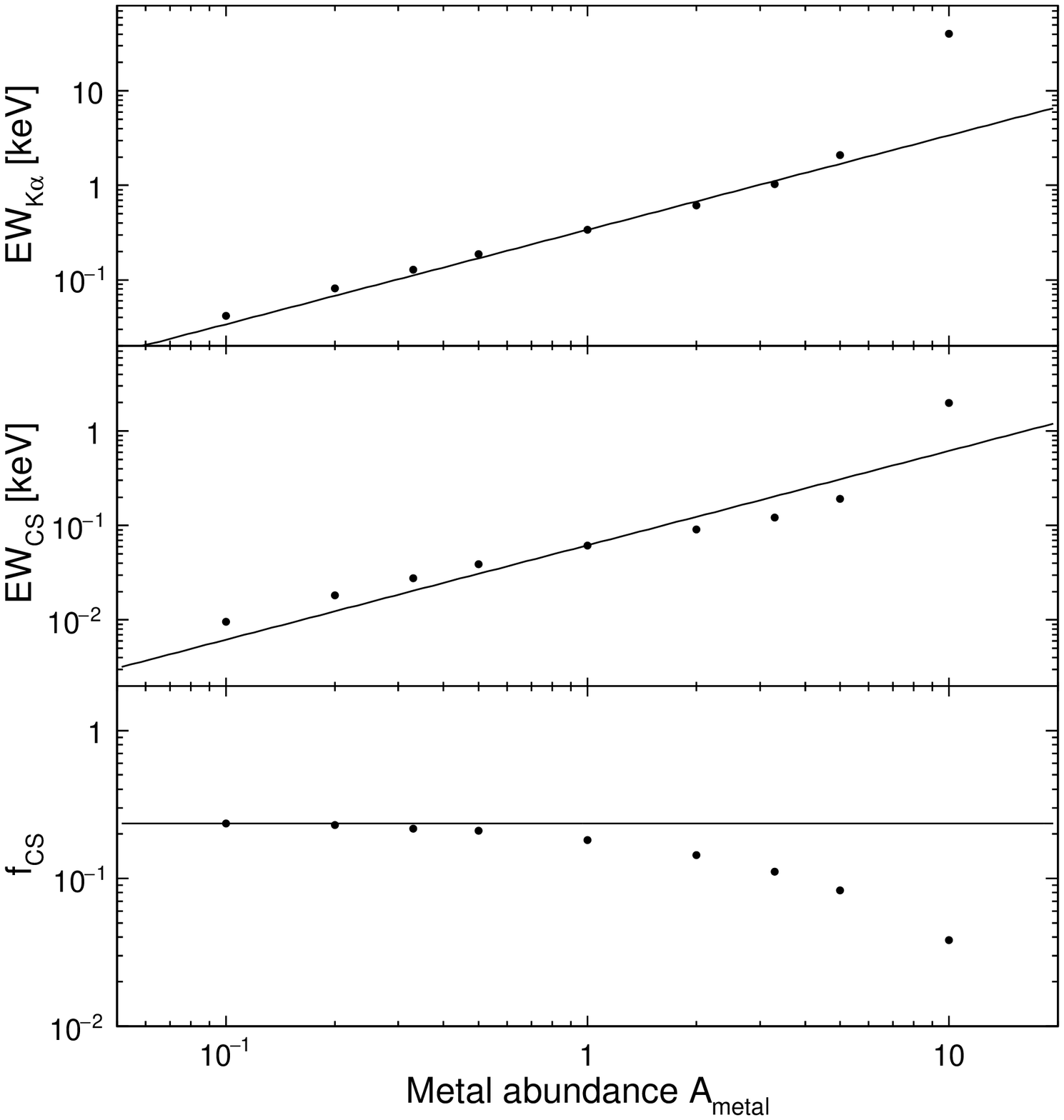}
\caption{Dependence of the iron line and its CS on the column density $N_\mathrm{H}$ (shown in the left panel) and on the metal abundance $A_\mathrm{metal}$ (right) of the spherical cloud . The data points show the three spectral quantities, equivalent width $\mathrm{EW_{K\alpha}}$ of the iron K$\alpha$ lines including their CS, equivalent width $\mathrm{EW_{CS}}$ of the CS, and the fraction $f_\mathrm{CS}$ of the CS as functions of $N_\mathrm{H}$ or of $A_\mathrm{metal}$. The error bars are due to statistical errors of Monte-Carlo simulations. We also superpose approximate functions given by Equation~(\ref{eq:relation_sphere}) in each panel as solid lines.
}
\label{fig:sphere_NH_MA_functions}
\end{center}
\end{figure*}

\begin{figure*}
\begin{center}
\includegraphics[width=8.0cm]{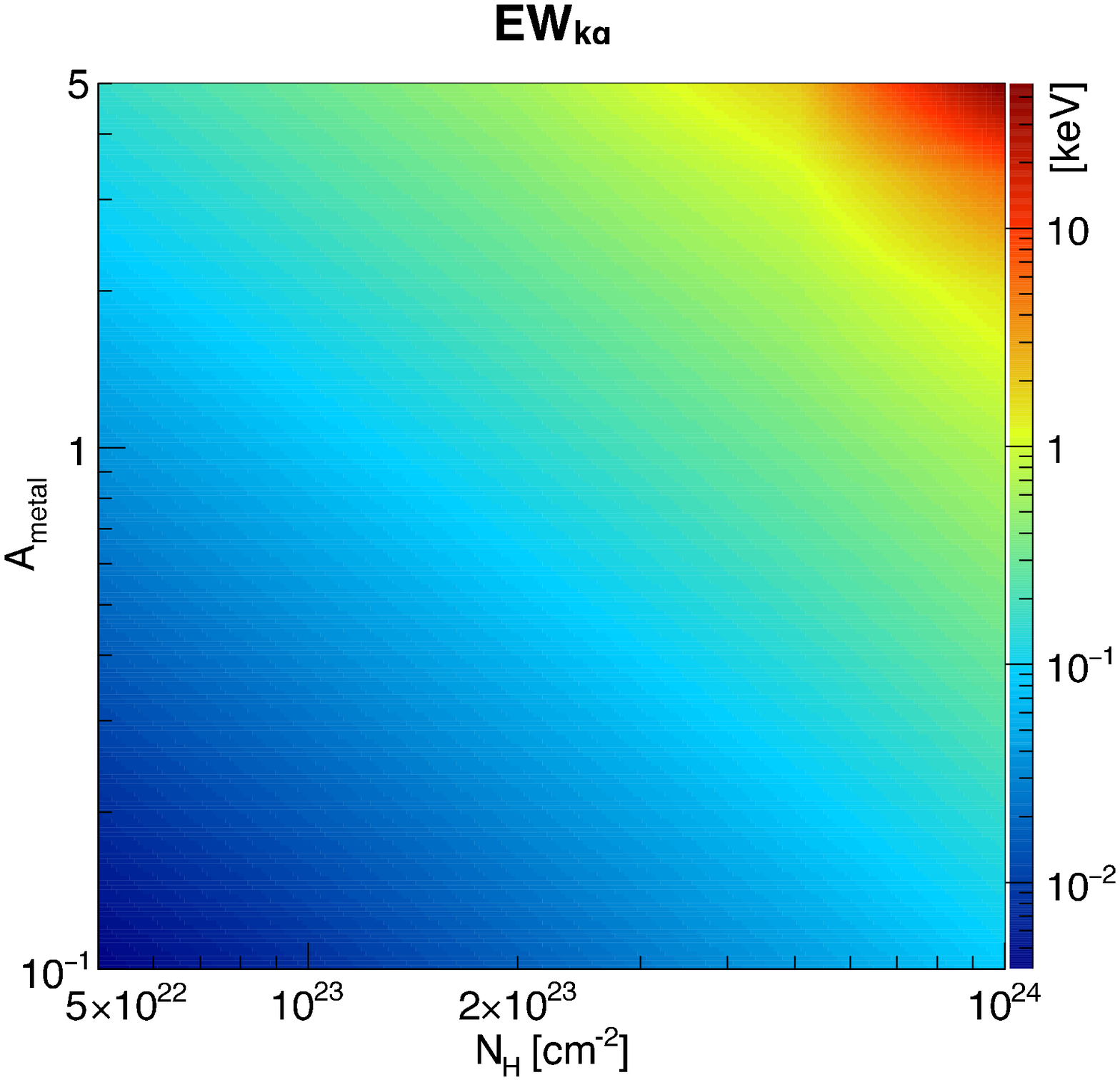}
\includegraphics[width=8.0cm]{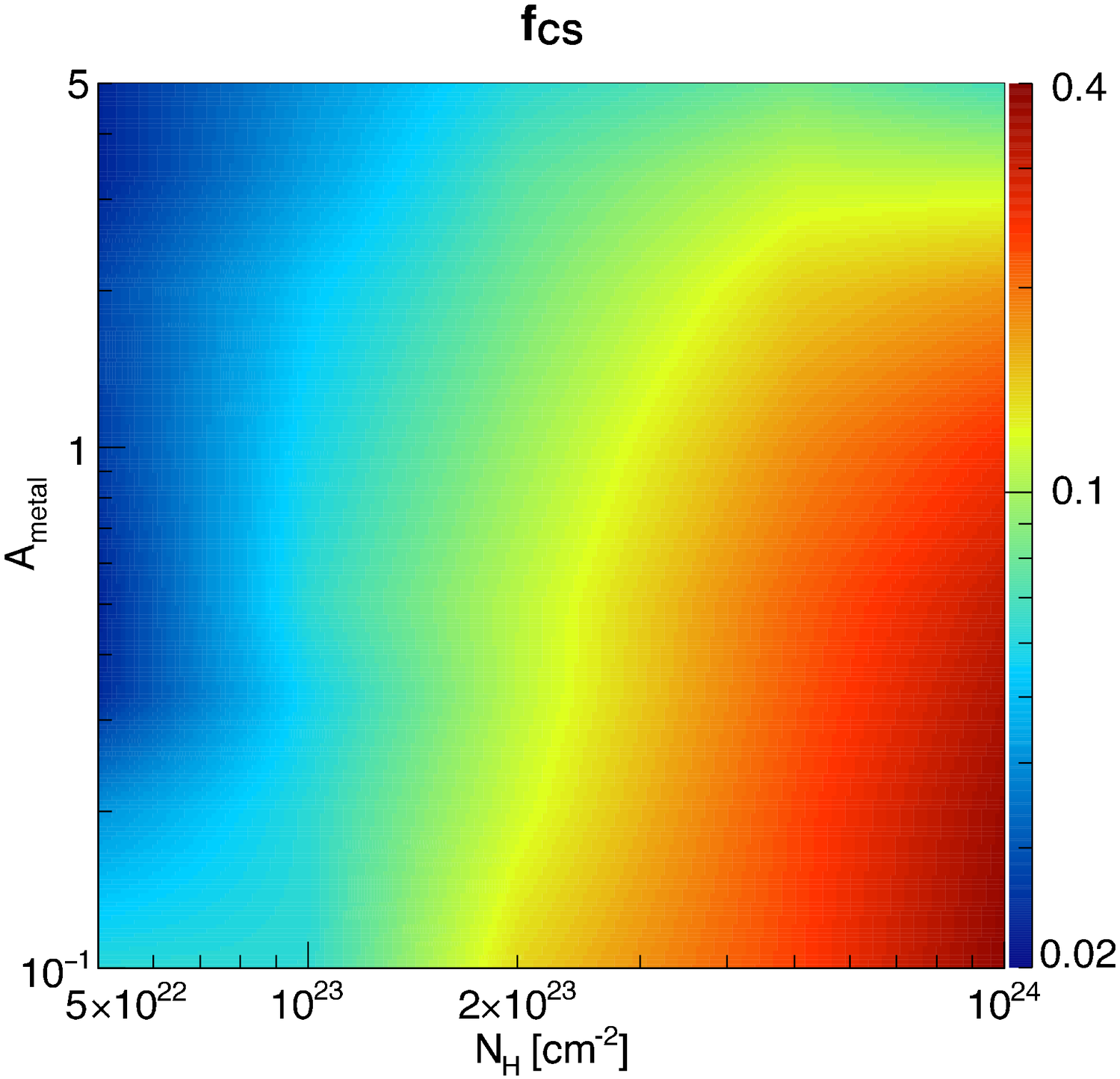}
\caption{The two-dimensional colour maps show the line equivalent width $\mathrm{EW_{K\alpha}}$ (in the left panel) and the fraction $f_\mathrm{CS}$ of the CS (in the right panel) as two-dimensional functions of the column density $N_\mathrm{H}$ and the metal abundance $A_\mathrm{metal}$.}
\label{fig:sphere_2D_MA_NH}
\end{center}
\end{figure*}

We calculate X-ray spectra emerging from different column densities ranging from $N_\mathrm{H}=2\times 10^{22}\;\mathrm{cm^{-2}}$ to $2\times 10^{24}\;\mathrm{cm^{-2}}$.
This range covers from a sufficiently optically thin value both for photoelectric absorption and for scattering up to a Thomson depth of 1.3, above which a spectrum shows too absorbed to evaluate the continuum.
The upper-left panel of Figure~\ref{fig:sphere_NH_MA_spectra} shows the calculated spectra in a broad band.
As $N_\mathrm{H}$ increases, absorption features---extinction at low energies and an iron K-shell absorption edge at 7.1 keV---become significant.
The iron line and its CS also become prominent with the increase of $N_\mathrm{H}$.
To see behaviour of the CS, we show in the upper-right panel of Figure~\ref{fig:sphere_NH_MA_spectra} the continuum-subtracted spectra around the iron line.
This figure shows how the absolute amount of the CS changes with $N_\mathrm{H}$.
At $N_\mathrm{H}<5\times 10^{22}\;\mathrm{cm^{-2}}$, the optical thickness of the cloud is not sufficient to produce the CS.
At $N_\mathrm{H}=1\times10^{23}\;\mathrm{cm^{-2}}$, the CS becomes visible, and its absolute amount increases to the peak at $N_\mathrm{H}=5\times10^{23}\;\mathrm{cm^{-2}}$.
Above this value of $N_\mathrm{H}$, the CS decreases with the column density since the absorption dominates over the generation of the CS.

We show three spectral properties defined in Section~\ref{subsec:analysis}, the EW of the whole iron K$\alpha$ line including the CS, the EW of the CS, and the fraction of the CS in the whole line, as functions of $N_\mathrm{H}$ in Figure~\ref{fig:sphere_NH_MA_functions} (the left panel).
Since $\mathrm{EW}_\mathrm{K\alpha}$ is a ratio of the generated iron line to the continuum, it is almost proportional to a probability of iron line generation, namely $N_\mathrm{H}\times A_\mathrm{metal}$.
The simulated data points well agree with a linear function of $N_\mathrm{H}$ as superposed in the figure.
Then, $\mathrm{EW}_\mathrm{CS}$ should be proportional to $N_\mathrm{H}{}^2\times A_\mathrm{metal}$ since the CS is a result of Compton scattering of the iron line, and therefore an additional factor of $N_\mathrm{H}$ is necessary.
The CS fraction $f_\mathrm{CS}$ should be proportional to $N_\mathrm{H}$ as it is considered as a probability of a line photon to be scattered.
The discussion above can be summarised as the following relations:
\begin{equation}
\label{eq:relation_sphere}
\begin{aligned}
\mathrm{EW_{K\alpha}} &\propto N_\mathrm{H}\times A_\mathrm{metal}, \\ 
\mathrm{EW_{CS}} &\propto N_\mathrm{H}{}^2\times A_\mathrm{metal}, \\
f_\mathrm{CS} &\propto N_\mathrm{H}.
\end{aligned}
\end{equation}

As well as the column density, the metal abundance $A_\mathrm{metal}$ is an important parameter that affects the spectrum via photoelectric absorption.
The bottom panels of Figure~\ref{fig:sphere_NH_MA_spectra} show spectra for different metal abundances in which $N_\mathrm{H}$ is fixed to $5\times 10^{23}\;\mathrm{cm^{-2}}$, and also corresponding continuum-subtracted spectra around the iron line.
In Figure~\ref{fig:sphere_NH_MA_functions} (the right panel), we show the three spectral quantities, $\mathrm{EW}_\mathrm{K\alpha}$, $\mathrm{EW}_\mathrm{CS}$, and $f_\mathrm{CS}$, as functions of $A_\mathrm{metal}$.
As described above, $\mathrm{EW}_\mathrm{K\alpha}$ and $\mathrm{EW}_\mathrm{CS}$ are proportional to $A_\mathrm{metal}$.
The slight decline of $\mathrm{EW}_\mathrm{CS}$ at high metal abundances is an effect of absorption.
The above discussion would also predict constant $f_\mathrm{CS}$ for different $A_\mathrm{metal}$, and it is true at low metal abundances, but it is actually a decreasing function of $A_\mathrm{metal}$ since absorption is more significant for the CS due to the longer trajectories of the scattered photons than those of the line photons which have not experienced scattering.

The column density is coupled with the metal abundance on the spectral features generated by photoelectric absorption.
As seen in the left column of Figure~\ref{fig:sphere_NH_MA_spectra}, the extinction by absorption seen in a broadband spectrum depends both upon the column density and the metal abundance, and both the contributions are almost indistinguishable in the broadband spectrum.
$\mathrm{EW_{K\alpha}}$ has a similar coupling problem since it is proportional to $N_\mathrm{H}\times A_\mathrm{metal}$.
These are due to a fact that the photoelectric effect is simply sensitive to amount of iron that makes extinction and line emissions (iron has the most significant contribution in this energy range).
The CS, however, has different behaviour between the column density and the metal abundance, as clearly seen in the right column of Figure~\ref{fig:sphere_NH_MA_spectra}.
Thus, it is a good idea to measure the CS in addition to the line since it is generated by Compton scattering, not by photoelectric effect to decouple them.
The CS fraction $f_\mathrm{CS}$ is therefore a suitable measure of $N_\mathrm{H}$.
Figure~\ref{fig:sphere_2D_MA_NH} shows two-dimensional dependence of $\mathrm{EW_{K\alpha}}$ and $f_\mathrm{CS}$ upon the two coupling properties on the photoelectric effect.
$f_\mathrm{CS}$ is almost independent from the metal abundance; moreover, the two quantities have opposite dependence on the metal abundance at high column densities, which makes the decoupling easier.

\subsection{Dependence on spectral slope of illuminating radiation}

\begin{figure*}
\begin{center}
\includegraphics[width=7.5cm]{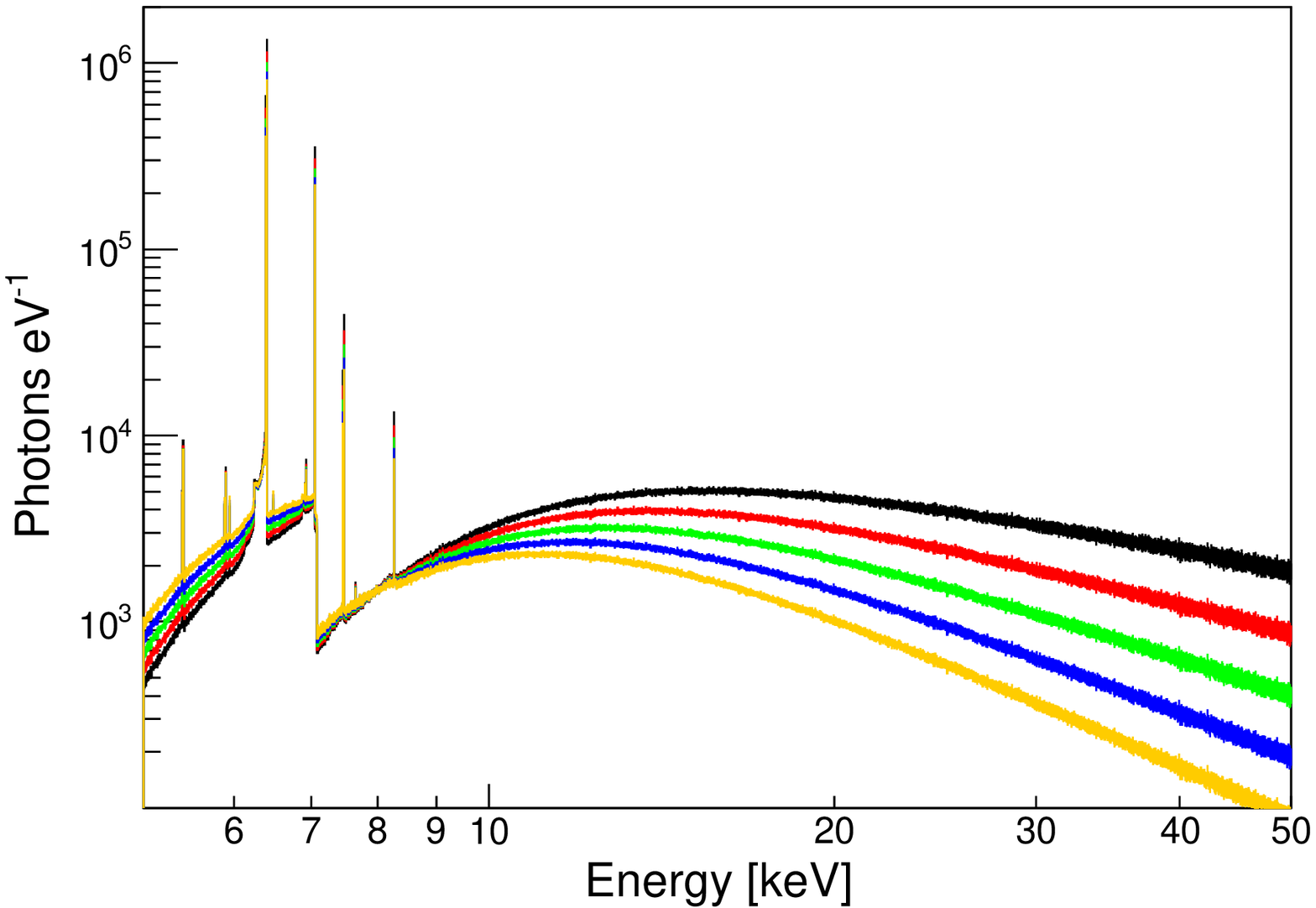}
\includegraphics[width=7.5cm]{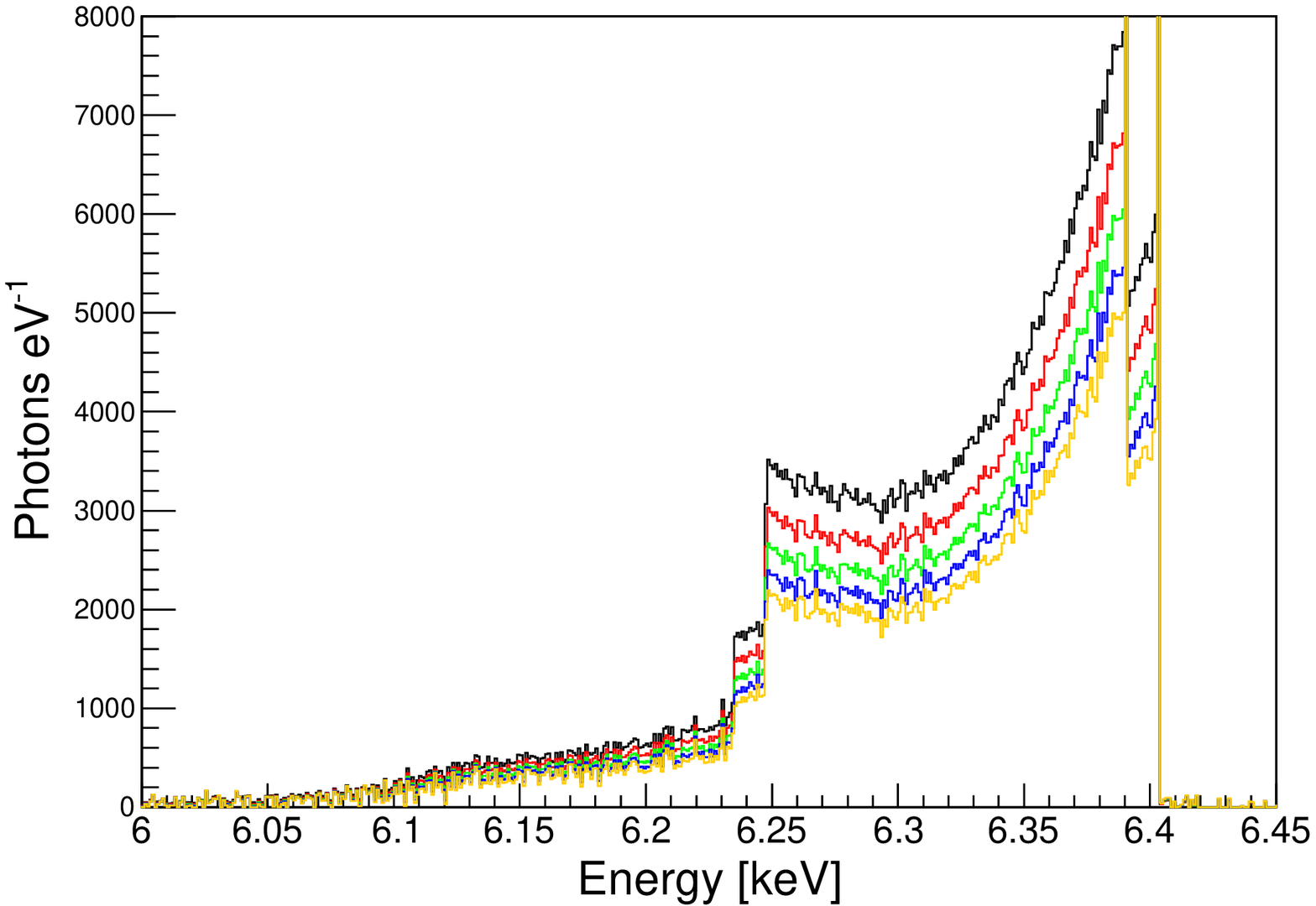} \\
\includegraphics[width=7.5cm]{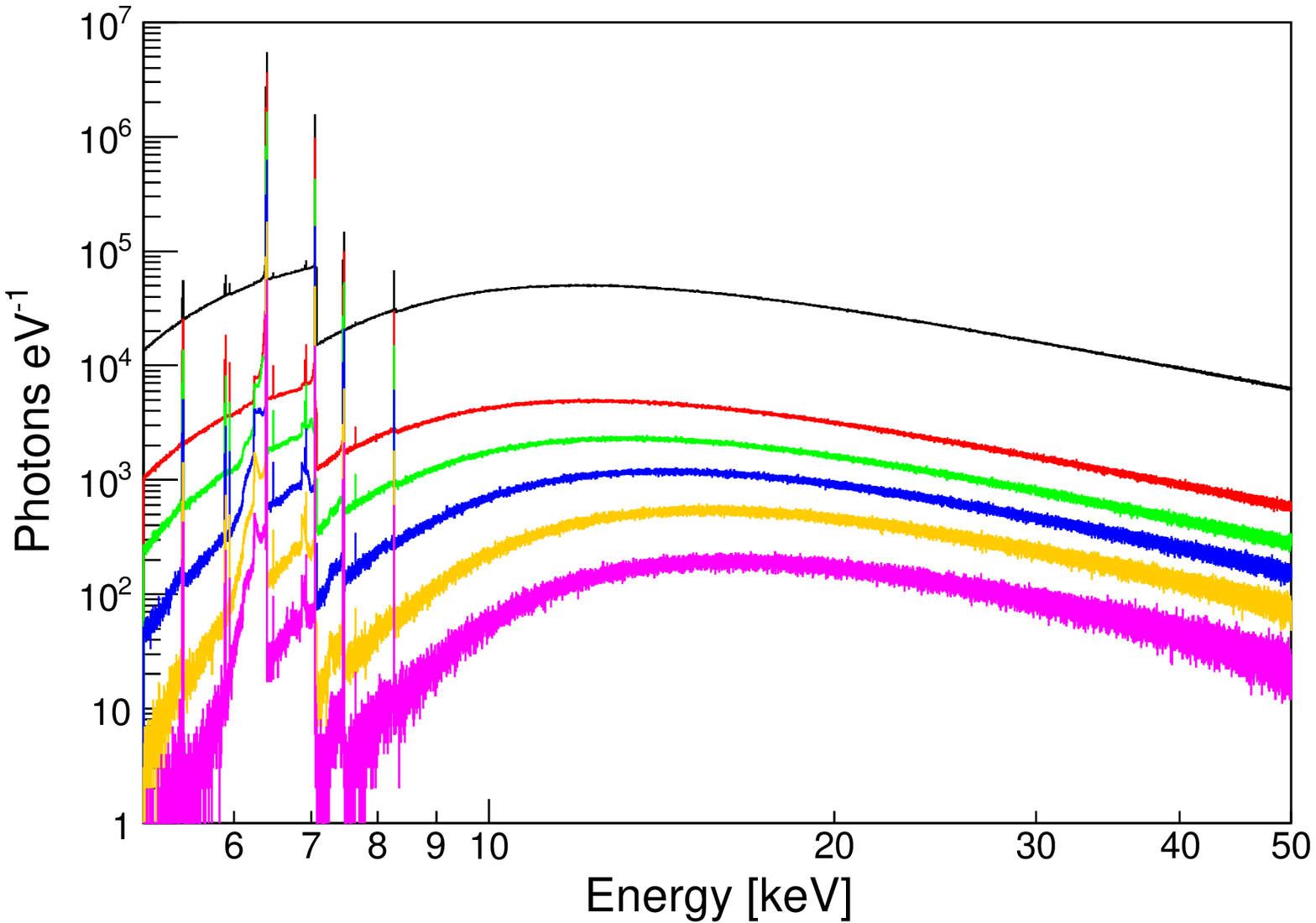}
\includegraphics[width=7.5cm]{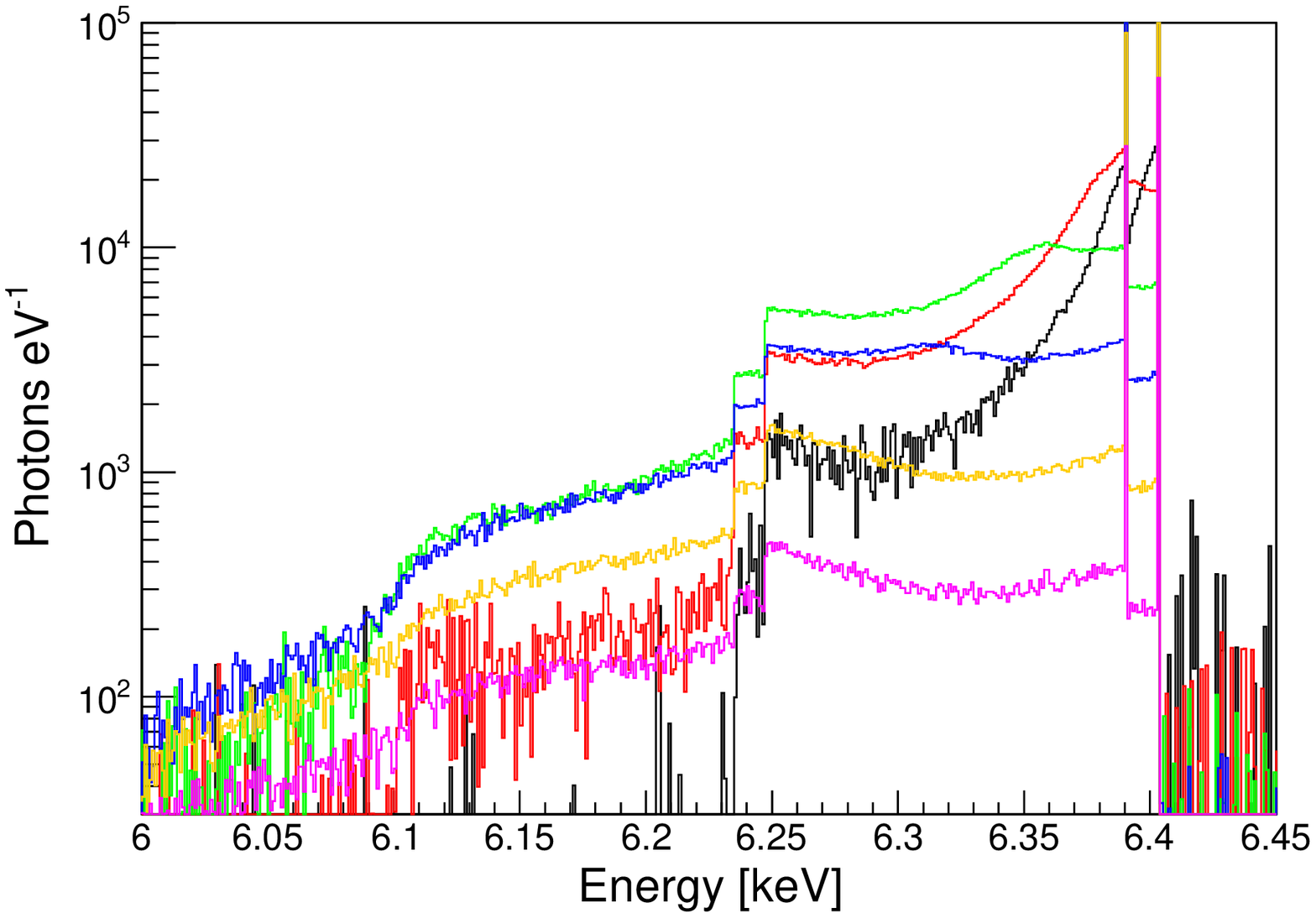}
\caption{X-ray spectra emerging from a spherical cloud with $N_\mathrm{H}=1\times 10^{24}\;\mathrm{cm^{-2}}$ and $A_\mathrm{metal}=1.0$ for different conditions of the X-ray illumination. Broadband spectra are shown in the left column, and spectra enlarged around the iron line in the right column where the underlaying continua are subtracted for clear comparison of the CS profile. In the upper panel, we apply different photon indices of the illuminating spectrum, $\Gamma=1.2$ (black), $1.6$ (red), $2.0$ (green), $2.4$ (blue), and $2.8$ (yellow).
In the bottom panels, we assume a short flare, and show spectra at different observation time: $0\le t<0.1$ (black), $0.1\le t<0.2$ (red), $0.5\le t<0.6$ (green), $1.0\le t<1.1$ (blue), $1.5\le t<1.6$ (yellow), and $2.0\le t<2.1$ (magenta). The origin of time is defined as the moment at which the direct photons reach to the observer. See text in detail.}
\label{fig:sphere_Gamma_time_spectra}
\end{center}
\end{figure*}

\begin{figure*}
\begin{center}
\includegraphics[width=7.75cm]{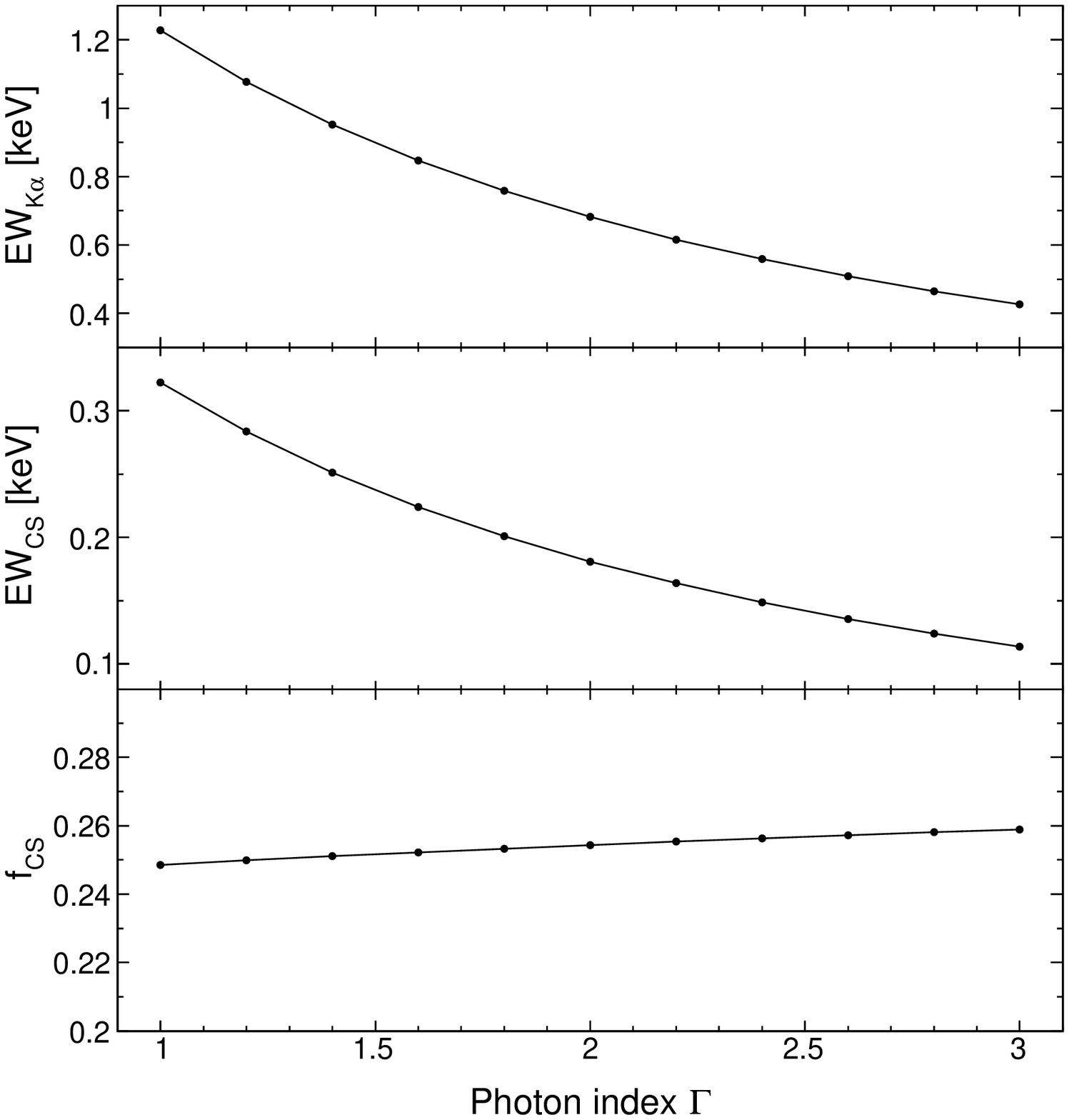}
\includegraphics[width=7.75cm]{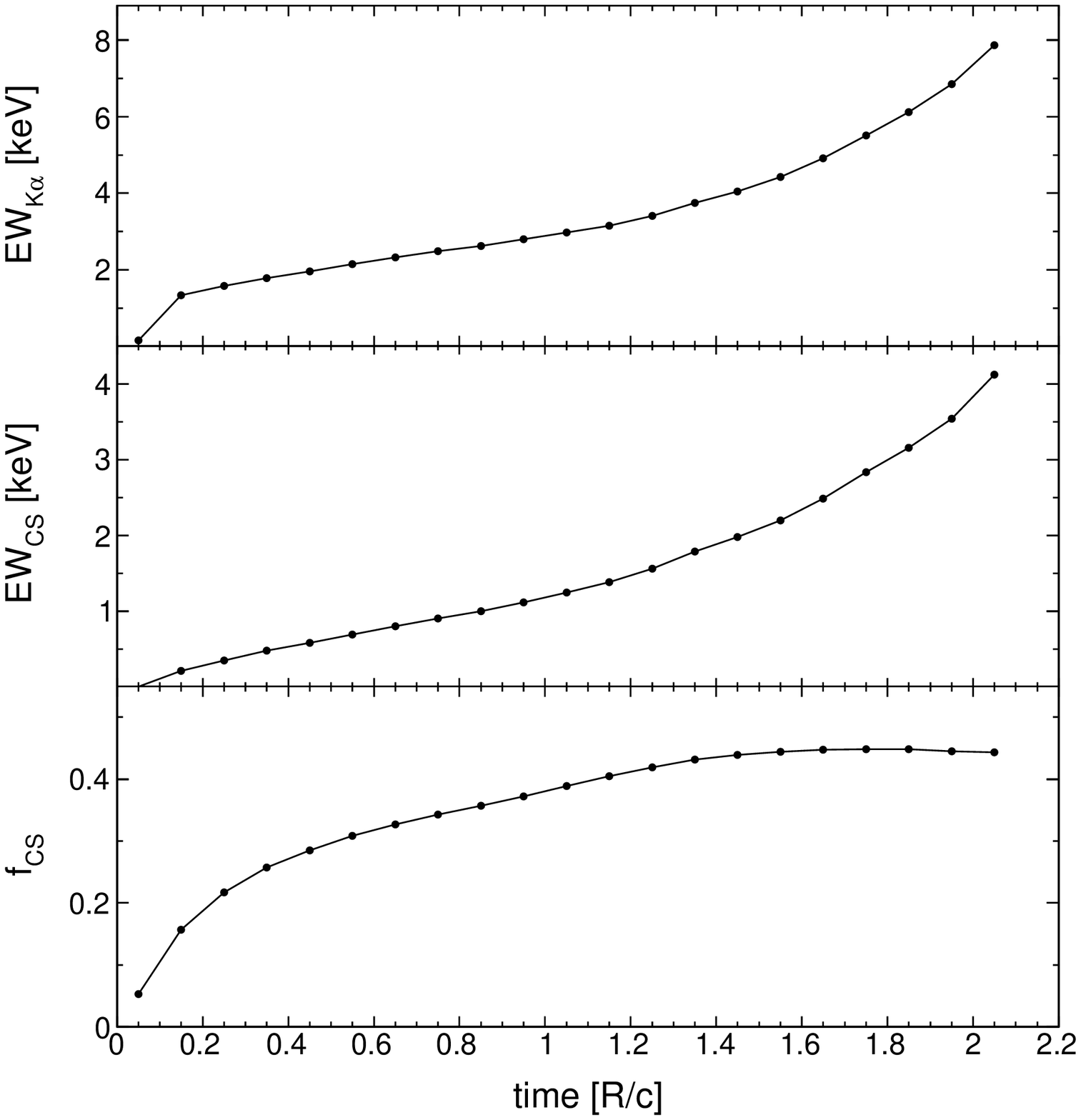}
\caption{Left: the data points show the three spectral quantities, equivalent width $\mathrm{EW_{K\alpha}}$ of the iron K$\alpha$ lines including their CS, equivalent width $\mathrm{EW_{CS}}$ of the CS, and the fraction $f_\mathrm{CS}$ of the CS as functions of the photon index $\Gamma$ of the illuminating spectrum.
Right: the time evolution of the three spectral quantities after a short flare are shown. The origin of time is defined as the moment at which the direct photons reach to the observer.}
\label{fig:sphere_Gamma_time_functions}
\end{center}
\end{figure*}

As the spectrum emerging from the cloud obviously depends on the intrinsic radiation, it is necessary to check dependence of the spectral quantities on properties of the illuminating source.
The upper panels of Figure~\ref{fig:sphere_Gamma_time_spectra} demonstrate how the spectrum changes with the photon index of the illuminating radiation for a cloud with $N_\mathrm{H}=1\times 10^{24}\;\mathrm{cm^{-2}}$ and $A_\mathrm{metal}=1.0$.
If two spectra have the same continuum levels at the 6.4-keV iron line, the harder spectrum supplies more photons above the K-shell edge at 7.1 keV to generate the iron line.
Therefore, $\mathrm{EW_{K\alpha}}$ increases with the hardness of the spectrum, as shown in the left panel of Figure~\ref{fig:sphere_Gamma_time_functions}.
This effect is well known and we have to measure (or assume) the spectral slope of the intrinsic radiation when we discuss line EW.

In contrast, we can exploit the CS as a spectral feature that does not depend on the intrinsic spectrum.
The top-right panel of Figure~\ref{fig:sphere_Gamma_time_spectra} shows the spectra around the iron line from which the continua are subtracted.
The shape of the CS does not change with the photon index $\Gamma$ since the iron line photons lose spectral information on the illuminating source.
Thus, $\mathrm{EW_{CS}}$ increases with the spectral hardness in the same way $\mathrm{EW_{K\alpha}}$ does, as shown in Figure~\ref{fig:sphere_Gamma_time_functions}, and more importantly, $f_\mathrm{CS}$ is almost perfectly constant with changing photon index.

\subsection{Time evolution of CS}

Since the fluorescence line and its CS are results of reprocessing of the intrinsic radiation, they must show a timing delay if the illuminating source has time variability.
This timing response can also be used for diagnostics of the cloud.
We calculate the time evolution of the spectrum if the time profile of the X-ray source is represented by a delta function.
Here we assume a cloud with $N_\mathrm{H}=1\times 10^{24}\;\mathrm{cm^{-2}}$ and $A_\mathrm{metal}=1.0$.
In astrophysics, this situation corresponds to an event such as a short bright flare of the central black hole.
In other words, the duration of the flare is assumed to be sufficiently short compared with the light crossing time of the cloud.

The calculated spectra observed at different time are shown in the bottom panels of Figure~\ref{fig:sphere_Gamma_time_spectra}.
The origin of time is defined as the moment at which the direct flare photons reach to the observer, and time $t$ is measured in units of $R/c$ ($R$ is the cloud radius).
The first (highest) spectrum drawn as a black solid line is a spectrum observed in $0\le t < 0.1$, and includes photons that directly propagate to the observer without any interactions.
After the direct photons pass the observer, or $t>0$, observed photons should be reprocessed ones by photoelectric effect or scattering inside the cloud, and the spectrum rapidly decays with time.

The CS profile shown in the bottom-right panel of Figure~\ref{fig:sphere_Gamma_time_spectra}
apparently evolves with time.
In $0\le t <0.1$, the CS has very hard spectrum, i.e., most of the photons in the CS have high energies.
A lower energy photon should have delay since it requires a larger scattering angle, and then means a longer path to reach the observer.
Thus, lack of low energy photons in a CS indicates the very initial phase of the flare.
Afterwards, the lower energy part increases with time and the profile approaches to a flat spectrum.

Time evolutions of the three spectral quantities are shown in the right panel of Figure~\ref{fig:sphere_Gamma_time_functions}.
$\mathrm{EW_{K\alpha}}$ rapidly evolves at the initial phase ($t<0.1$) far beyond an EW value of 1 keV, and then gradually increases with time.
$\mathrm{EW_{CS}}$ has similar behaviour but has slight delay owing to an additional scattering.
These increases of the EWs are the result of the faster decay of the continuum at the iron line energy compared with the line itself, which is generated by photons above the K-edge.
$f_\mathrm{CS}$ increases with time, being almost saturated after $t=1.5$.
The combination of $\mathrm{EW_{K\alpha}}$ and $f_\mathrm{CS}$ is useful for estimating time elapsed since the flare, as already discussed by \citet{Odaka:2011}.
\citet{Sunyaev:1998} also focused on the time evolution of the line EW for a longer time range.

\section{Slab Geometry}
\label{sec:slab}

In this section, we investigate X-ray reflection including a CS from a slab geometry.
As drawn in the bottom panel of Figure~\ref{fig:geometry}, an X-ray emitter is placed on the top of a disc, and the direction of generated photons are set to be vertically downward, i.e., $(0,\;0,\;-1)$.
This setup simulates a situation in which a plane wave illuminates the slab that has infinite horizontal dimensions since the disc radius in the simulation is set to be sufficiently large so that effects of the edge of the disc are negligible.
This geometry is the simplest geometry to allow us to see angular dependence of the CS.
This slab geometry has three parameters: a vertically measured hydrogen column density $N_\mathrm{H}$ (see the figure), a metal abundance $A_\mathrm{metal}$, and a viewing angle $\theta$ (or inclination angle).
A simulation with $2\times 10^{9}$ primary photons was conducted for each parameter pair of ($N_\mathrm{H}$, $A_\mathrm{metal}$).

\subsection{Angular dependence}
\label{subsec:slab_angular}

\begin{figure*}
\begin{center}
\includegraphics[width=7.5cm]{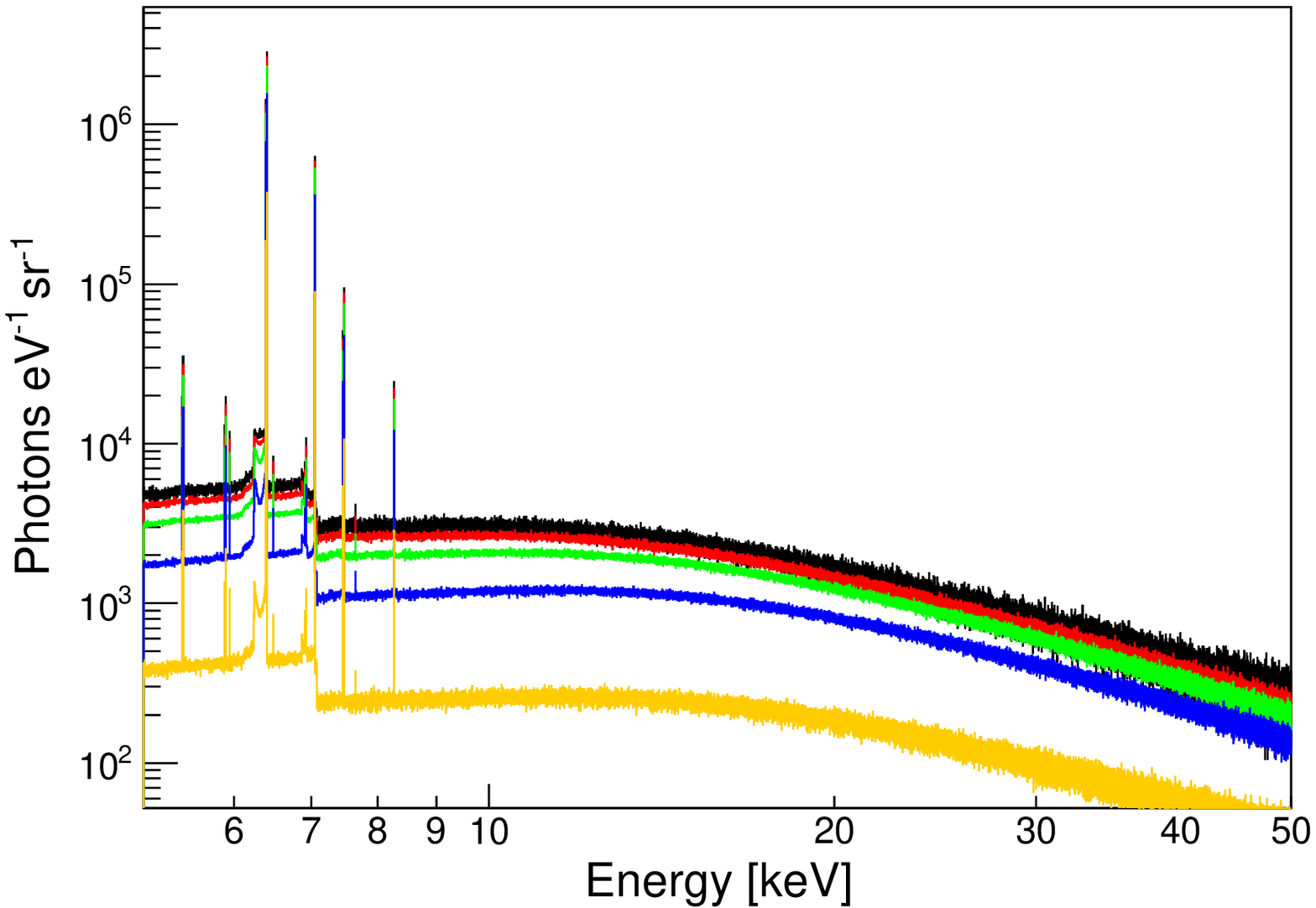}
\includegraphics[width=7.5cm]{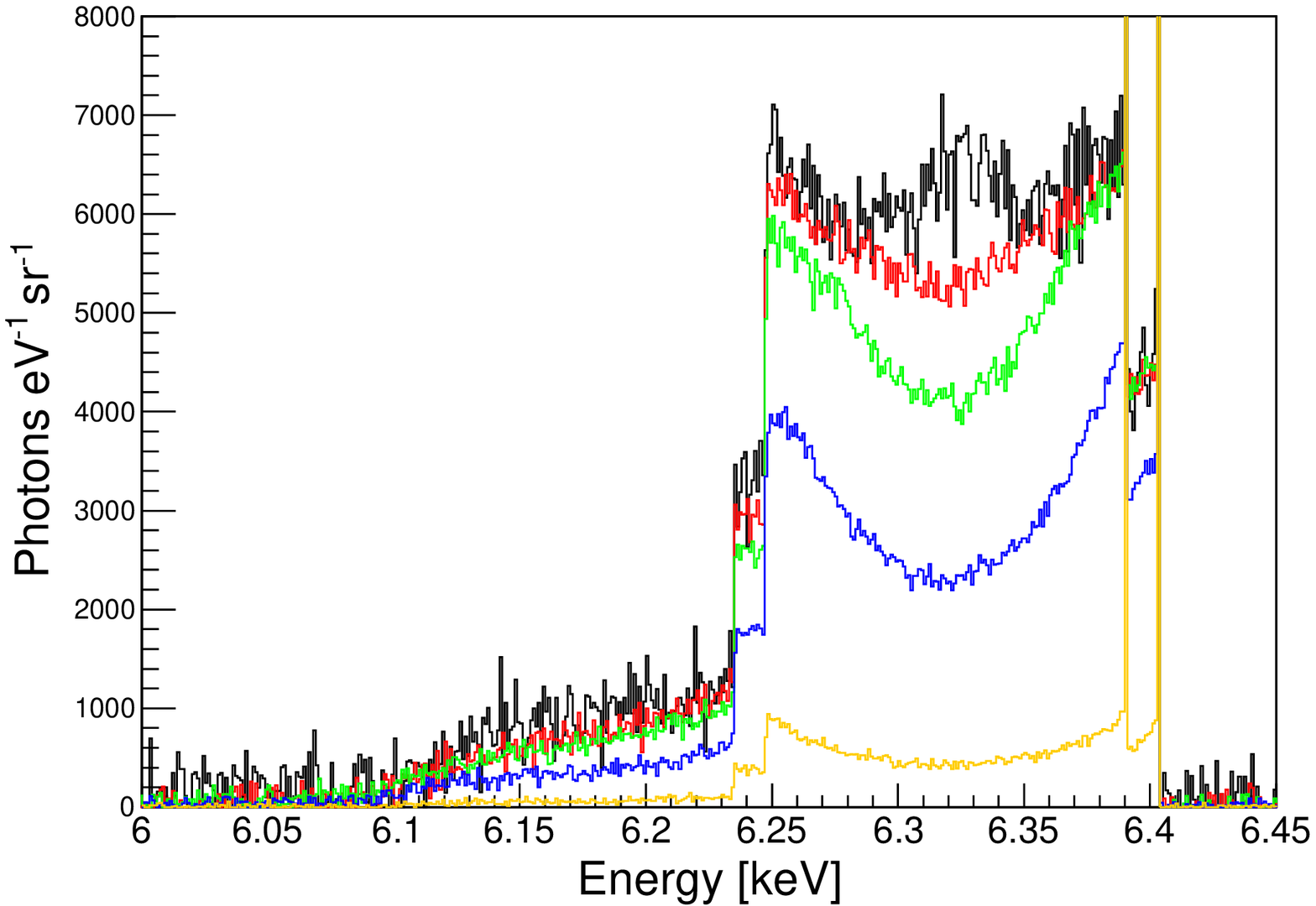}
\caption{X-ray spectra emerging from a slab with $N_\mathrm{H}=1\times 10^{24}\;\mathrm{cm^{-2}}$ and $A_\mathrm{metal}=1.0$ observed from different viewing angles of $0^\circ \le \theta < 10^\circ$ (black), $20^\circ \le \theta < 30^\circ$ (red), $40^\circ \le \theta < 50^\circ$ (green), $60^\circ \le \theta < 70^\circ$ (blue), and $80^\circ \le \theta < 90^\circ$ (yellow).
Broadband spectra are shown in the left column, and spectra enlarged around the iron line in the right column where the underlaying continua are subtracted for clear comparison of the CS profile. The normalisations of the spectra are corrected so that each spectrum covers unit solid angle.}
\label{fig:slab_angle_NH1e24_spectra}
\end{center}
\end{figure*}

\begin{figure}
\begin{center}
\includegraphics[width=7.75cm]{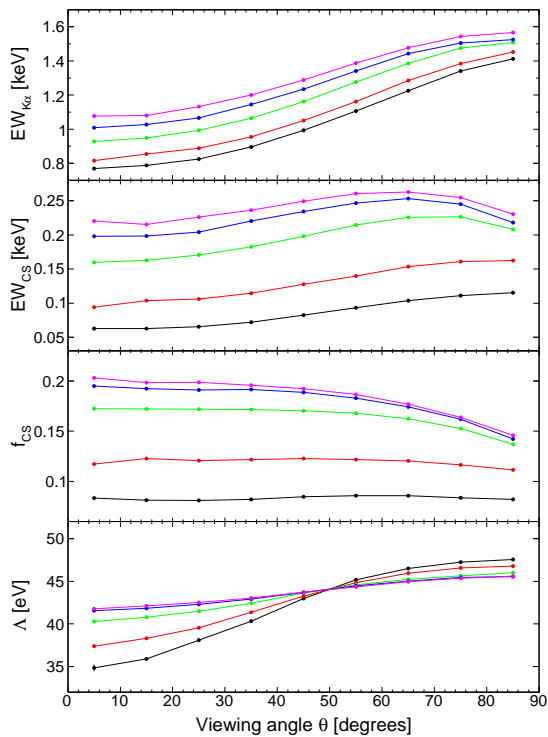}
\caption{The data points show the four spectral quantities, equivalent width $\mathrm{EW_{K\alpha}}$ of the iron K$\alpha$ lines including their CS, equivalent width $\mathrm{EW_{CS}}$ of the CS, the fraction $f_\mathrm{CS}$ of the CS, and the deviation $\Lambda$ of the CS distribution as functions of the viewing angle $\theta$. Simulation results with different column densities $N_\mathrm{H}=1\times 10^{23}\;\mathrm{cm^{-2}}$ (black), $2\times 10^{23}\;\mathrm{cm^{-2}}$ (red), $5\times 10^{23}\;\mathrm{cm^{-2}}$ (green), $1\times 10^{24}\;\mathrm{cm^{-2}}$ (blue), and $1\times 10^{25}\;\mathrm{cm^{-2}}$ (magenta) are shown.}
\label{fig:slab_angle_functions}
\end{center}
\end{figure}

We extract several spectra that have different viewing angles from each simulation output with ($N_\mathrm{H}$, $A_\mathrm{metal}$).
In spite of the similarity in the spectral shapes in a broad band, the profiles of the CSs display very different shapes as shown in Figure~\ref{fig:slab_angle_NH1e24_spectra}.
Particularly the face-on reflection (observed from $0^\circ \le \theta < 10^\circ$) has a peak at 6.32 keV which corresponds to a scattering angle of $90^\circ$.
The enhancement of scattering at $90^\circ$ for the face-on spectrum is a result of the large anisotropy of the slab geometry.
Viewed from $\theta=0^\circ$, a photon to be scattered at $90^\circ$ is allowed to run along a horizontal path, which can have significantly large effective optical depth (yielding high probability of scattering).
More analytical treatment is described in Appendix \S\ref{sec:analytical_profile}.
Spectra from other directions, on the other hand, do not have a peak at this energy, or rather they are suppressed here.

The enhancement due to the horizontal path becomes less significant if the vertical optical depth $\tau$ approaches to unity simply because the contribution to scattering of matter beyond $\tau\sim 1$ is not effective and so the anisotropy vanishes.
This effect on the CS profile can be evaluated by using the ``deviation'' $\Lambda$ of the spectral distribution of the CS which we introduced in \S\ref{subsec:analysis}.
The bottom panel of Figure~\ref{fig:slab_angle_functions} shows angular dependence of $\Lambda$ for different $N_\mathrm{H}$.
The case of $N_\mathrm{H}=1\times 10^{23}\;\mathrm{cm^{-2}}$ shows an effective enhancement at 6.32 keV, which is implied by a small value of $\Lambda$.
As $N_\mathrm{H}$ increases, this effect gets suppressed, and its behaviour almost converges at $N_\mathrm{H}\sim 1\times 10^{24}\;\mathrm{cm^{-2}}$.

Figure~\ref{fig:slab_angle_functions} also shows angular dependence for several different column densities of the three spectral quantities which we discussed in \S\ref{sec:sphere} for the spherical cloud.
Although the EWs of the line and the CS do not significantly change with the viewing angle, these angular dependences can be understood as follows.
$\mathrm{EW_{K\alpha}}$ increases with $\theta$ since the continuum, which dominantly consists of scattered photons of the incident primaries, has dipolar distribution while the line is more isotropic.
$\mathrm{EW_{CS}}$ has angular dependence similar to that of $\mathrm{EW_{K\alpha}}$ at small $N_\mathrm{H}$, but if $N_\mathrm{H}$ becomes larger, the absorption effectively reduces the CS photons.
This effect is readily seen in the plot of $f_\mathrm{CS}$ for large $N_\mathrm{H}$.

\subsection{Dependence on column density and metal abundance}

\begin{figure*}
\begin{center}
\includegraphics[width=8.0cm]{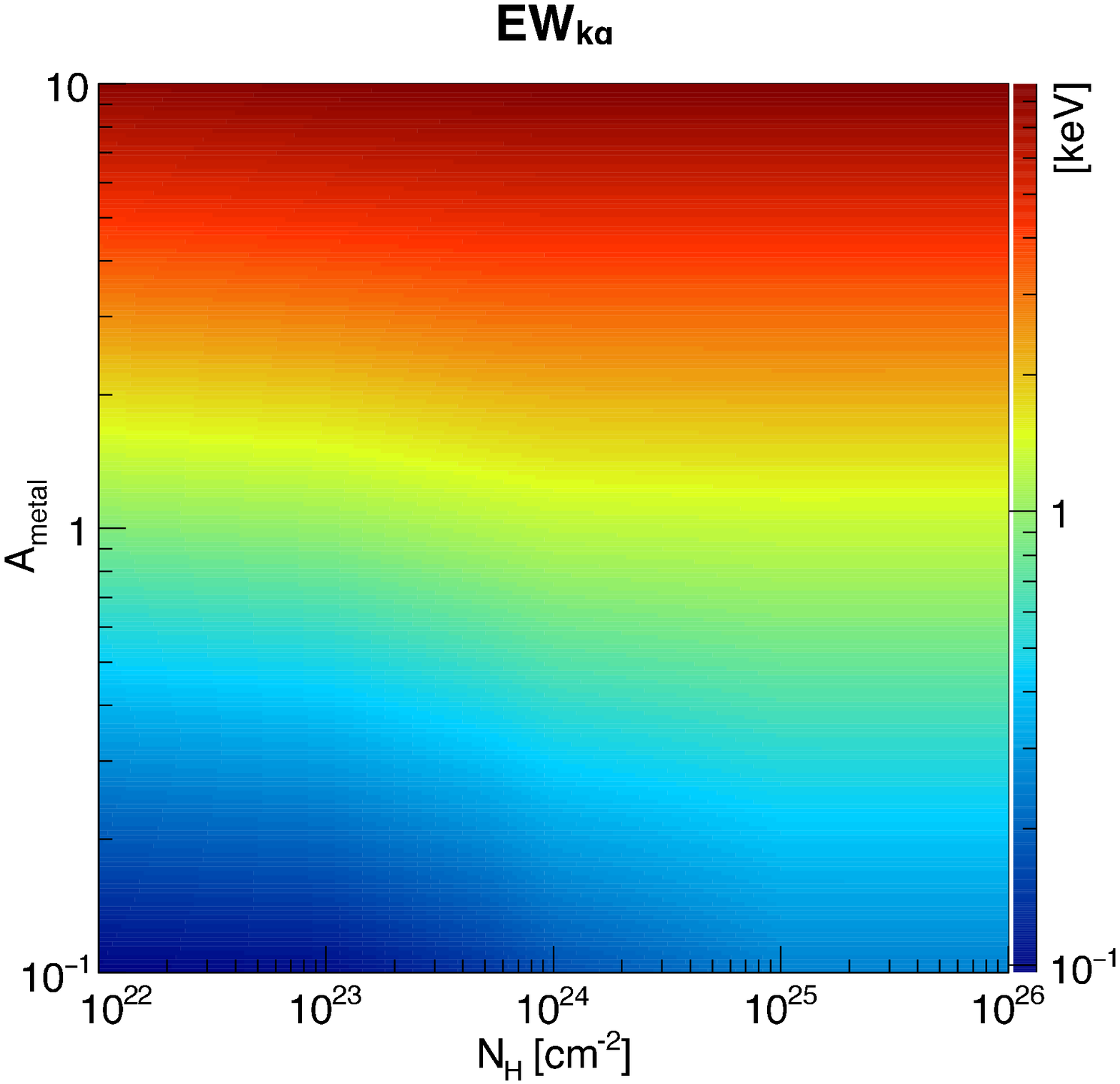}
\includegraphics[width=8.0cm]{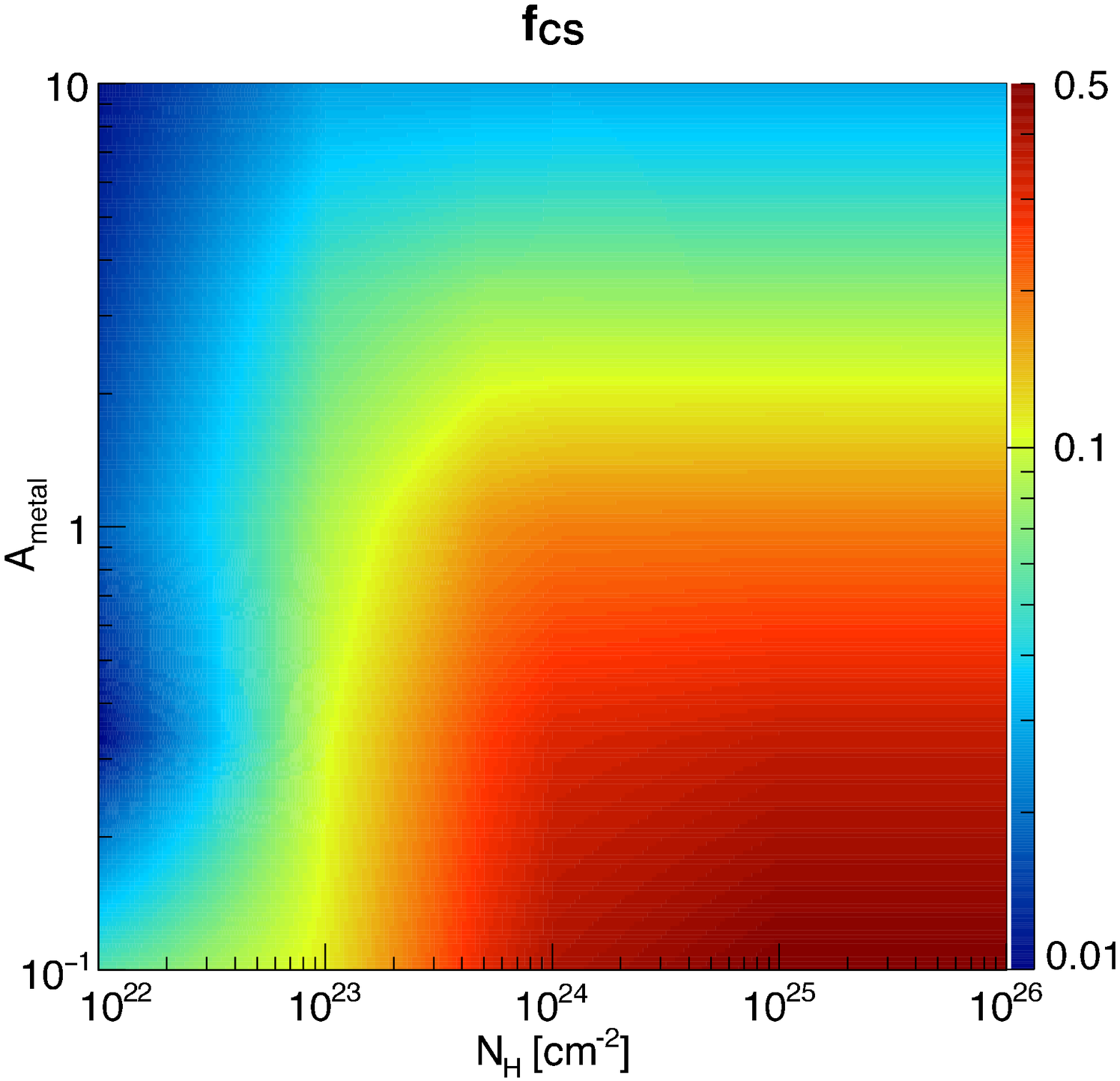}
\caption{The two-dimensional colour maps show the line equivalent width $\mathrm{EW_{K\alpha}}$ (the left panel) and the fraction $f_\mathrm{CS}$ of the CS (the right panel) as two-dimensional functions of $N_\mathrm{H}$ and $A_\mathrm{metal}$. A slab geometry is assumed and the spectra are extracted for viewing angles of $40^\circ$--$50^\circ$.}
\label{fig:slab_2D_MA_NH}
\end{center}
\end{figure*}

We again investigate the dependence on the column density and the metal abundance for the slab geometry.
Figure~\ref{fig:slab_2D_MA_NH} shows $\mathrm{EW_{K\alpha}}$ and $f_\mathrm{CS}$ as two-dimensional functions of $N_\mathrm{H}$ and $A_\mathrm{metal}$.
Here we limit the viewing angle to a range of $40^\circ \le \theta < 50^\circ$, while the overall behaviour is very similar among the different viewing angles.
By comparison with the case of the spherical geometry shown in Figure~\ref{fig:sphere_2D_MA_NH}, the dependence is apparently different.
This difference is caused by absence of the direct (transmitted) component rather than a geometrical effect.
Recalling the discussion in \S\ref{subsec:sphere_dependence_NH_MA}, we can derive relations as
\begin{equation}
\label{eq:relation_slab}
\begin{aligned}
\mathrm{EW_{K\alpha}} &\propto A_\mathrm{metal}, \\ 
\mathrm{EW_{CS}} &\propto N_\mathrm{H}\times A_\mathrm{metal}, \\
f_\mathrm{CS} &\propto N_\mathrm{H},
\end{aligned}
\end{equation}
since the reprocessed continuum is also proportional to $N_\mathrm{H}$, dropping a factor of $N_\mathrm{H}$ from them.
As already discussed, the CS suffers from absorption due to a longer photon path, and therefore the approximate functions of $\mathrm{EW_{CS}}$ and $f_\mathrm{CS}$ are no longer appropriate when absorption is significant, or $N_\mathrm{H}\times A_\mathrm{metal}$ is large.

If the slab is highly Compton-thick, or $N_\mathrm{H}>10^{25}\;\mathrm{cm^{-2}}$, the spectral quantities of the reprocessed component become constant on the column density.
This is quite natural since the X-ray photons are not able to penetrate into the deep region of the slab, and therefore the effective column density is saturated around $N_\mathrm{H}\sim 10^{25}\;\mathrm{cm^{-2}}$.
Thus, it is difficult to probe such a highly Compton-thick object by using the iron fluorescence, and the hard X-ray band above 20 keV, where absorption has insignificant impact on the spectrum, provides more important observational means.

The CS, or $f_\mathrm{CS}$ if quantitatively, is useful to decouple the two parameters, $N_\mathrm{H}$ and $A_\mathrm{metal}$, on the absorption.
$f_\mathrm{CS}$ is almost independent of the metal abundance unless the absorption is significant.
Even if the absorption has great impact at high metal abundances, $f_\mathrm{CS}$ has  anti-correlation in the case of Compton-thin ($N_\mathrm{H} < 1\times 10^{24}$), which allows us to distinguish the two parameters coupling on the absorption.

\section{Effects of physical state of target electrons}
\label{sec:electron_states}

\begin{figure}
\begin{center}
\includegraphics[width=8.0cm]{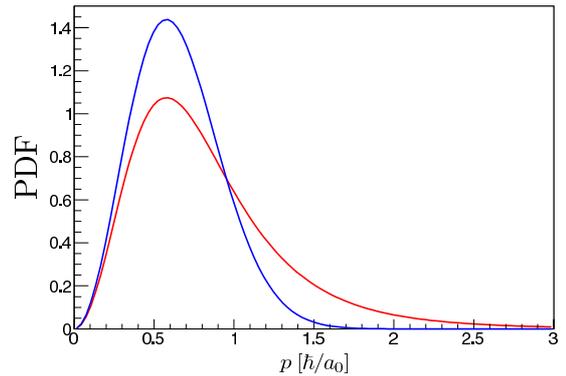}
\caption{Comparison between the probability distribution function (PDF) of momentum of electrons bound to atomic hydrogen (red) and that of free electrons with a temperature of $kT=4.53\;\mathrm{eV}$ (blue). Both functions give the same peak position at $p=2.15\;\mathrm{keV}/c=(1/\sqrt{3})\hbar a^{-1}$.}
\label{fig:momentum_distribution}
\end{center}
\end{figure}

\begin{figure}
\begin{center}
\includegraphics[width=7.5cm]{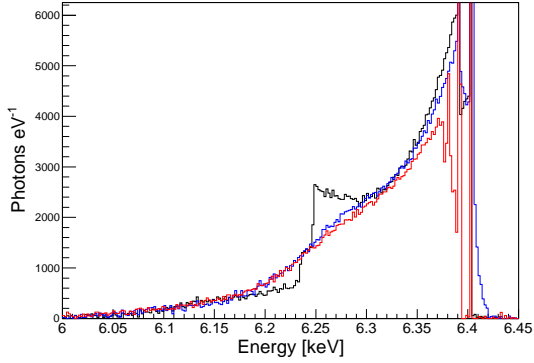}
\caption{Spectra of the CSs emerging from a sphere with $N_\mathrm{H}=1.0\times 10^{24}\;\mathrm{cm^{-2}}$ and $A_\mathrm{metal}=1.0$ for different electron states. The underlying continua are subtracted to clearly show the shapes of the CSs. We calculate three conditions in which electrons are (1) free at rest (shown in black), (2) bound to atoms of hydrogen or helium (red), and (3) free in a plasma with an electron temperature of $kT=5\;\mathrm{eV}$ (blue).}
\label{fig:estate_spectra_zoom_sphere}
\end{center}
\end{figure}

\begin{figure}
\begin{center}
\includegraphics[width=7.5cm]{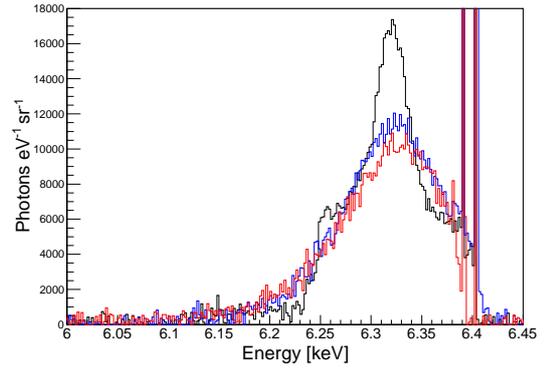} \\
\includegraphics[width=7.5cm]{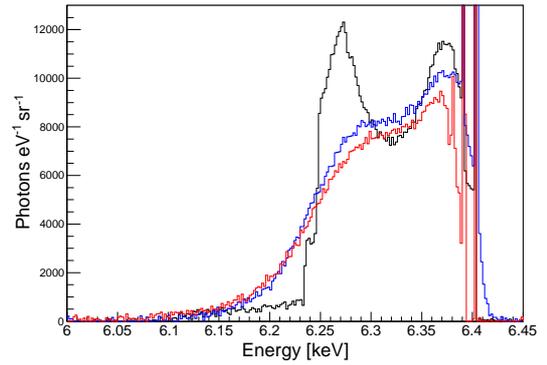} \\
\includegraphics[width=7.5cm]{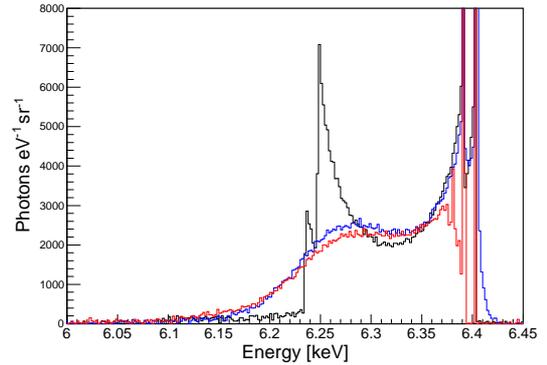}
\caption{The same as Figure~\ref{fig:estate_spectra_zoom_sphere} but for a slab geometry with $N_\mathrm{H}=1.0\times 10^{23}\;\mathrm{cm^{-2}}$ and $A_\mathrm{metal}=1.0$. The spectra are seem from different viewing angles of $0^\circ \le \theta < 10^\circ$ (top), $40^\circ \le \theta < 50^\circ$ (middle), and $80^\circ \le \theta < 90^\circ$ (bottom). The simulation for each electron state was performed with $2.4\times 10^{10}$ photons.}
\label{fig:estate_spectra_zoom_slab}
\end{center}
\end{figure}

\begin{table}
\caption{Spectral quantities for different electron states}
\begin{center}
\begin{tabular}{cccc}
\hline\hline
Electron state & $\mathrm{EW_{K\alpha}}$ & $\mathrm{EW_{CS}}$ & $f_\mathrm{CS}$ \\
 & [keV] & [keV] & \\ 
\hline
Free at rest & 0.682 & 0.181 & 0.254 \\
Bound & 0.684 & 0.157 & 0.218 \\
Free ($kT=5\;\mathrm{eV}$) & 0.669 & 0.171 & 0.244 \\
\hline
\end{tabular}
\end{center}
\label{table:estate_quantities}
\textbf{Notes:} These values are extracted from simulations of a spherical cloud with $N_\mathrm{H}=1.0\times 10^{24}\;\mathrm{cm^{-2}}$ and $A_\mathrm{metal}=1.0$.
\end{table}%

In a general astrophysical environment, electrons may have various properties that affect the CS; e.g., electrons can have momentum due to random or bulk motion of the matter, or due to thermal motion.
Even if the scattering medium is completely at rest, an electron bound to an atom or a molecule has momentum in the quantum system which causes the Doppler effect.
These physical states of electrons make observable effects on properties of the CS, and therefore they should be considered in spectral analysis concerning the CS.
For instance, velocity dispersion smears the CS profile by the Doppler effect \citep{Yaqoob:2010}.
In this section, we focus on the case of cold matter completely at rest in which all electrons are bound to atoms.
This physical setup can be applicable to cases of a molecular cloud, an AGN torus, or dense gas surrounding an X-ray binary.
Although it is, of course, highly possible that some of the electrons are ionised or have velocity, we here demonstrate how the CS is modified purely by the electron binding effects.
We also show the case of free electrons that have moderate temperature in a plasma as reference.

\subsection{Basic consideration on momentum distributions}

It is helpful to think about the distribution of momentum of the target electrons before we see results of the Monte-Carlo simulations.
As physics implementation of \texttt{MONACO} being described in \S\ref{subsec:physics}, our simulation code is capable of treating effects of bulk and random motion of a target electron which includes thermal motion specified by an electron temperature, as well as the code has full implementation of the scattering process by a bound electron, which can be divided into the three channels---Rayleigh scattering, Raman scattering, and Compton scattering.
The calculation procedure for these physical processes is precise but is quite complicated to make an ``intuitive'' understanding of the results.
So we briefly explain a simple background behind the modification of the CS profile in a framework of the Doppler effect and the momentum distribution of the target electrons.

If an electron initially has a momentum $\bm{p}_e$ and so an energy $E_e=\sqrt{(m_e c^2)^2+(p_e c)^2}$, the energy of the scattered photon by this electron, which would be given by Equation~(\ref{eq:compton}) for an electron at rest, is modified as
\begin{equation}\label{eq:compton_motion}
h\nu_1 = \dfrac{h\nu_0}{1+\dfrac{h\nu_0}{E_e}(1-\cos\theta)} \left(1+\dfrac{\bm{p}_e\cdot (\bm{k}_1-\bm{k}_0)}{E_e E_0} \right),
\end{equation}
where $\bm{k}_0$ and $\bm{k}_1$ denote momenta of the photon before and after the scattering, respectively.
Using this relation, we can roughly say that a typical amount of energy change caused by the initial momentum of the electron is given by a factor of $p_e c/m_e c^2$.
Considering the ground state of atomic hydrogen, the averaged momentum is given by $p_e=\hbar/a=3.7\;\mathrm{keV}/c$, and therefore the typical energy shift due to the atomic binding is $\Delta E\sim 6.4\;\mathrm{keV}\times{p_ec/m_ec^2}=46\;\mathrm{eV}$.
This level of the energy shift will be obviously visible by a high-resolution microcalorimeter.
It is also worth nothing that this value is much larger than the electron binding energy of an atomic hydrogen, 13.6 eV.

Thermal motion of electrons can also be a cause of the Doppler effect appearing in the CS spectrum.
We consider here a plasma with a moderate temperature that results in smearing of the CS profile at a degree comparable to that due to the hydrogen binding.
The wave function in momentum space of the ground state of hydrogen atom is given by
\begin{equation}
\psi(p)=\frac{1}{\pi}\left(\dfrac{2a}{\hbar}\right)^\frac{3}{2}\frac{1}{(1+a^2p^2/\hbar^2)^2},
\end{equation}
and then this function gives the most probable value at $p=(1/\sqrt{3})\hbar a^{-1}$.
The Maxwell--Boltzmann distribution for a temperature of $kT$ ($k$ is the Boltzmann constant) gives the most probable value at $p=\sqrt{2m_e kT}$.
Figure~\ref{fig:momentum_distribution} shows comparison between the momentum distribution of the electrons bound to atomic hydrogens and that sampled from the Maxwellian distribution at $kT=4.53\;\mathrm{eV}$ which gives its peak (the most probable value) at the same value of $p=2.15\;\mathrm{keV}/c=(1/\sqrt{3})\hbar a^{-1}$.
The momentum distribution of bound electrons extends to higher value while the Maxwellian distribution has exponential decay.
Taking this difference of the distribution shapes into account, we adopt a slightly higher temperature, $kT=5\;\mathrm{eV}$ as a reference in the following calculations.

\subsection{Monte-Carlo results}

We performed simulations that have a geometrical setup of a spherical cloud identical to what is described in \S\ref{sec:sphere} but have different states of electrons.
We assume two different environments: cold matter and plasma at $kT=5\;\mathrm{eV}$.
In the cold matter setup, we assume matter with zero dynamical velocity in which all electrons responsible for scattering are bound to atomic hydrogens or heliums.
Note that electrons associated with metals have negligible contribution to scattering because of their small abundances. 
We incorporate all physics described in \S\ref{subsec:physics_bound_electron} into the simulations so that the modification of the cross sections and the differential cross sections of the scattering processes, which result in smeared CS profiles, is taken into account.
In the plasma case, a target electron is sampled from the Maxwell--Boltzmann distribution and the Doppler effect is treated in the framework of Lorentz transformation, as described in \S\ref{subsec:physics_free_electron}.

Figure~\ref{fig:estate_spectra_zoom_sphere} shows the CS spectra, where the underlying continua are subtracted, from a spherical cloud for the different states of electrons.
This figure demonstrates how the CS profile is smeared by the electron velocity due to atomic binding or thermal motion.
While the case of electrons at rest displays a sharp edge at 6.24 keV corresponding to the back scattering (scattering angle of $180^\circ$), this edge structure is blurred by the Doppler broadening in both conditions of the bound electrons and of the thermal electrons.
It is also seen that the blurred CS profiles of the both cases are remarkably similar because of the comparable momentum distribution.

Theoretically, there are small differences in the produced spectrum between the thermal electrons and the bound electrons.
The thermal motion produces a high energy tail (blue-shifted) above the line energy since a small fraction of the thermal electrons gives energy to photons via scattering.
The spectrum by the bound electrons, on the other hand, has line features produced via Raman scattering closely below the main line energy, which are never seen in the case of free electrons.
The energy shifts of these line features by Raman scattering from the main lines (Fe K$\alpha_1$ and K$\alpha_2$) correspond to excited levels of atomic hydrogen and helium.
But all these minor differences are visible only in a spectrum of extremely high quality.
Thus, it is difficult to distinguish the electron binding effect from the thermal blurring at an electron temperature $\sim 5\;\mathrm{eV}$.
This temperature corresponds to a dynamically random velocity at $\sqrt{3kT/m_e}\sim 1600\;\mathrm{km\;s^{-1}}$.

We also calculate spectra reflected from a slab geometry which is described in \S\ref{sec:slab} for the same electron conditions of motion.
We assume $N_\mathrm{H}=1.0\times 10^{23}\;\mathrm{cm^{-2}}$.
The results are shown in Figure~\ref{fig:estate_spectra_zoom_slab} for different viewing angles, face-on, middle ($\sim 45^\circ$), and edge-on.
The spectral features seen in the spectra for electrons at rest are all smeared by the electron motions.
The results from the bound electron resembles the results from thermal electrons as we see in the case of the spherical cloud.

Finally, we compare the EWs of the iron line and the CS and the fraction of the CS for the different electron states, and those values are shown in Table~\ref{table:estate_quantities}.
The values in this table are taken from simulations of a sphere with $N_\mathrm{H}=1.0\times 10^{24}\;\mathrm{cm^{-2}}$ and $A_\mathrm{metal}=1.0$.
The process of the iron line generation is identical among the three cases; thus, $\mathrm{EW_{K\alpha}}$ should have the same value among them.
But $\mathrm{EW_{K\alpha}}$ in the case of free electrons at 5 eV have a slightly smaller value (98\%) simply because the iron line and its CS are thermally broadened, those photons resulting in being outside of the energy range for integration.
The properties related to Compton scattering, $\mathrm{EW_{CS}}$ and $f_\mathrm{CS}$, are apparently influenced by the electron state, being reduced from values in the case of free electrons at rest if the electrons have finite temperature or are bounded to atoms.
One reason for this reduction is again the smearing of the CS profile which gets photons outside the integration range.
More importantly, if the electrons are bound to atomic hydrogens, $\mathrm{EW_{CS}}$ and $f_\mathrm{CS}$ get reduced due to the suppression of Compton scattering.
This reduction rate of 86\% for bound electrons, which can be found in Table~\ref{table:estate_quantities}, is understood by the suppression of the cross section of Compton scattering, as shown in Table~\ref{table:cross_section}.

\section{Discussion of observational data analysis}
\label{sec:observation}

\begin{figure}
\begin{center}
\includegraphics[width=8.5cm]{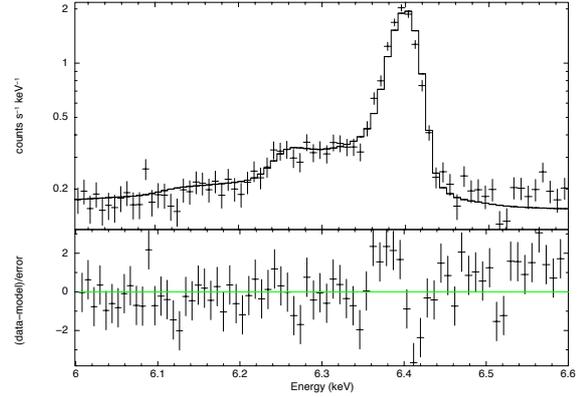}
\caption{Spectrum of GX 301$-$2 obtained with \textit{Chandra}-HETG. The data points are extracted from HEG$+1$ order. The table model generated by our simulations fitted to the data is shown in a solid line (see text).}
\label{fig:GX301-2_spectrum}
\end{center}
\end{figure}

\begin{figure*}
\begin{center}
\includegraphics[width=8.0cm]{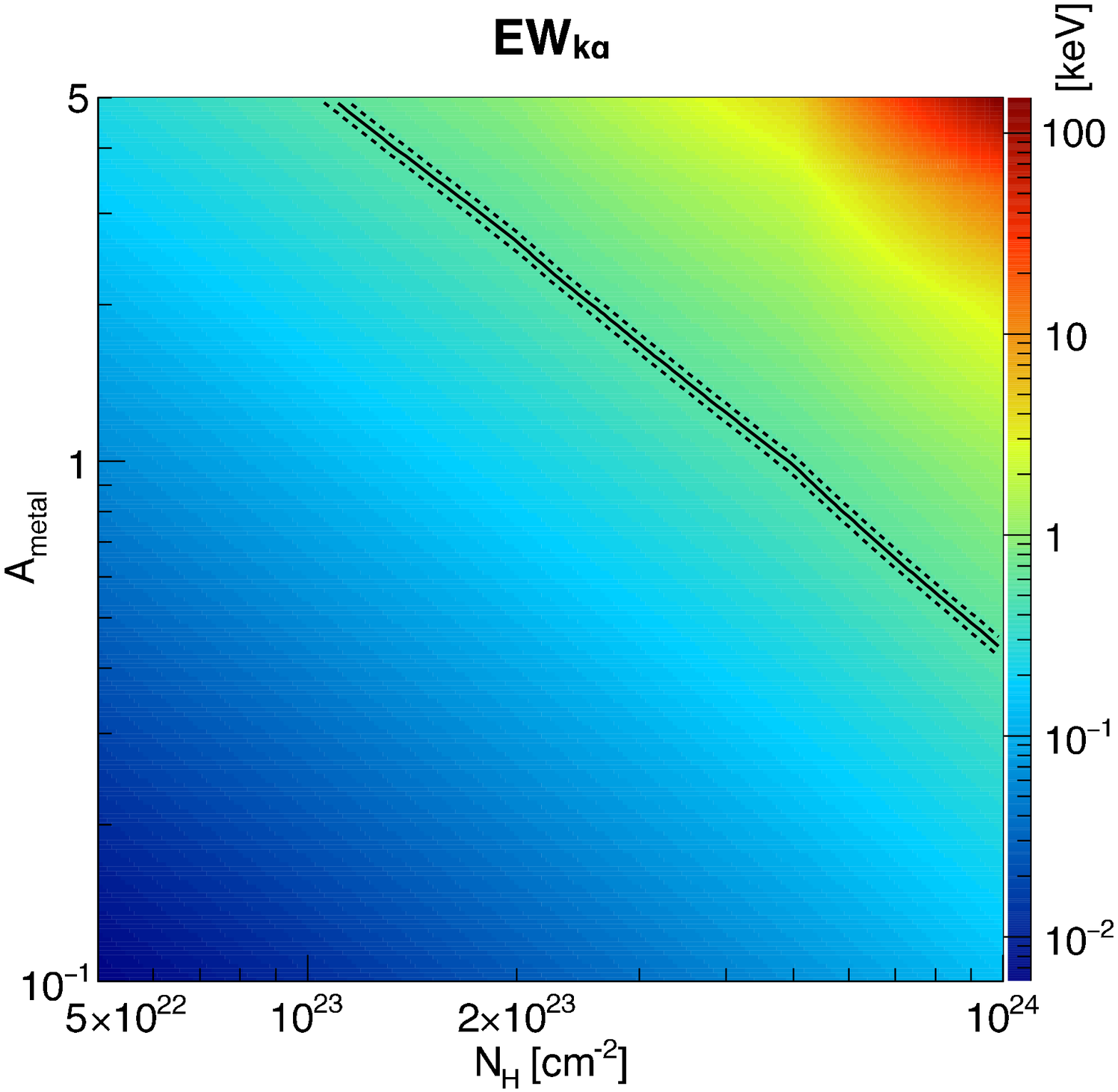}
\includegraphics[width=8.0cm]{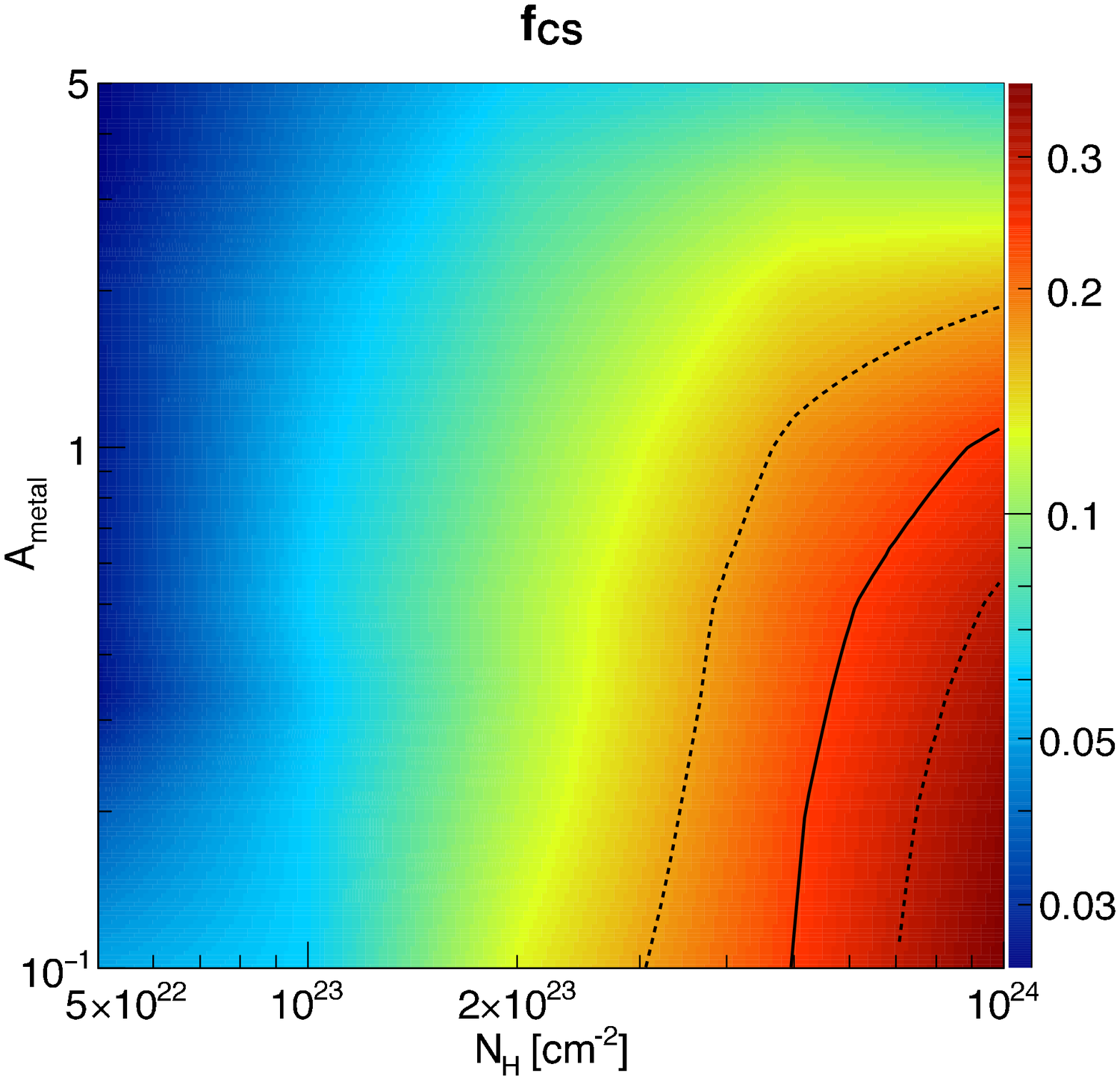}
\caption{The same as \ref{fig:sphere_2D_MA_NH} but that the illuminating spectrum has a photon index $\Gamma=1.0$. The values obtained by the data analysis of GX 301$-$2 are drawn in solid lines, and 1$\sigma$ errors in dashed lines.}
\label{fig:sphere_2D_MA_NH_pl1}
\end{center}
\end{figure*}

\begin{figure}
\begin{center}
\includegraphics[width=7.5cm]{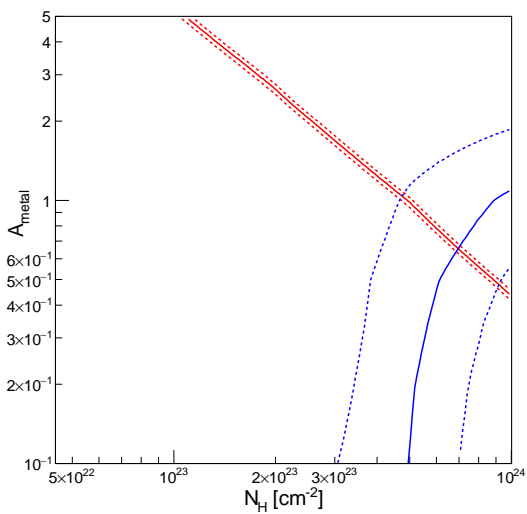}
\caption{Values of $N_\mathrm{H}$ and $A_\mathrm{metal}$ estimated by $\mathrm{EW_{K\alpha}}$ (red) and $f_\mathrm{CS}$ (blue) are drawn in the 2-dimensional parameter space.}
\label{fig:EWk_fCS_GX301-2}
\end{center}
\end{figure}

\begin{figure}
\begin{center}
\includegraphics[width=7.5cm]{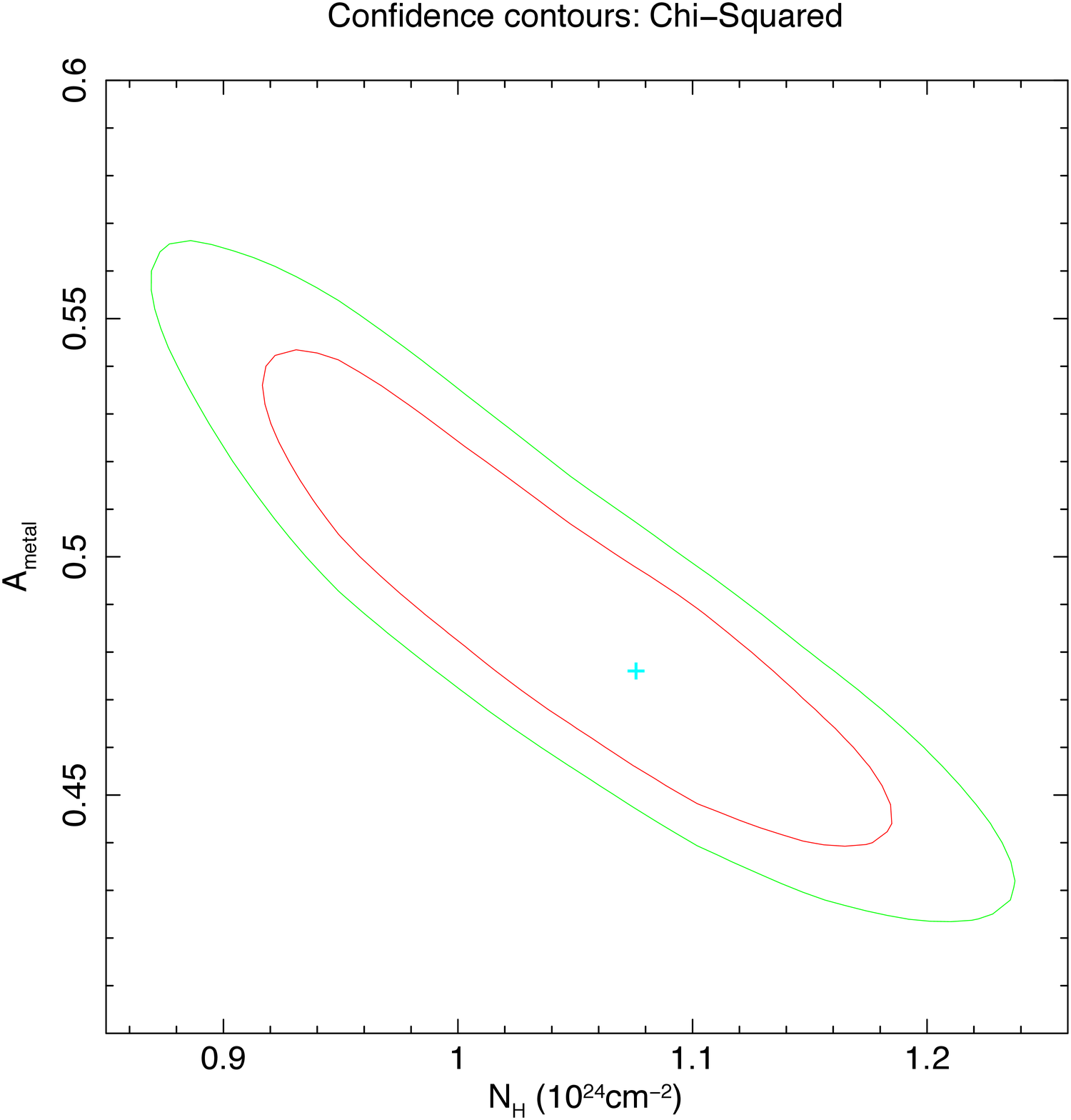}
\caption{Error contours of the fitting result at $1\sigma$ (red) and $2\sigma$ (green) levels in the two-dimensional space of $N_\mathrm{H}$ and $A_\mathrm{metal}$.}
\label{fig:contour}
\end{center}
\end{figure}



In the previous sections, we have investigated how the CS depends on spatial and temporal parameters of reflectors.
As we discuss in \S\ref{subsec:sphere_dependence_NH_MA}, column density and metal abundance, which would be coupled if we only used photoelectric absorption features and fluorescence lines, can be determined by using CS.
We also see that spectral slope of an illuminating source affects this analysis since $\mathrm{EW_{K\alpha}}$ is dependent on the photon index $\Gamma$.
Therefore, we can constrain the three parameters, $N_\mathrm{H}$, $A_\mathrm{metal}$, and $\Gamma$, by adding the CS in the spectral analysis to conventional analysis that exploits photoelectric absorption and fluorescence lines.
In this section, we demonstrate a simple way to constrain these parameters from an observational data which shows a CS.

We use a spectrum of HMXB GX 301$-$2 obtained with a grating spectrometer HETG onboard \textit{Chandra}.
The spectrum is extracted from data of Obs.\ ID 2733, which is one of data used in analysis by \citet{Watanabe:2003}, by using \texttt{CIAO} 4.8 and \texttt{CALDB} 4.7.0.
We use only the $+1$ order spectrum of HEG since the $-1$ order spectrum has a slightly complicated response around the iron line.
The obtained spectrum is shown in Figure~\ref{fig:GX301-2_spectrum}, displaying a clear feature of the CS.

We first determine the spectral quantities, $\mathrm{EW_{K\alpha}}$ and $f_\mathrm{CS}$.
For simplicity, we adopt an approximation that a spectrum can be obtained by dividing a count rate histogram by an effective area as a function of energy, which is called auxiliary response file or ARF in a standard data analysis of X-ray astronomy.
This approximation neglects broadening of the spectral response of the detector due to its finite energy resolution, being valid if the energy resolution were excellent.
Thus, this method will be suitable for microcalorimeters with excellent resolutions.
The spectrum is fitted by a power law in energy ranges of 5.4--5.69 keV and of 6.45--6.7 keV and a Gaussian line in 6.37--6.43 keV to determine the continuum and the line flux.
The intensity of the CS is calculated as an integral over 6.086--6.43 keV from which the line and continuum are subtracted.
We thus obtained $\mathrm{EW_{K\alpha}}=0.544\pm 0.022\;\mathrm{keV}$ and $f_\mathrm{CS}=0.236 \pm 0.063$ by characterising the iron K complex.

These values can be used to constrain the parameters of GX 301$-$2.
The object is an HMXB that has a bright accreting neutron star embedded in dense stellar wind from its donor star.
This astrophysical situation should be well described by the spherical model built in Section~\ref{sec:sphere}.
Figure~\ref{fig:sphere_2D_MA_NH_pl1} shows $\mathrm{EW_{K\alpha}}$ and $f_\mathrm{CS}$ as two-dimensional functions of $N_\mathrm{H}$, $A_\mathrm{metal}$, also showing constraint by the measured values.
These functions are the same as shown in Figure~\ref{fig:sphere_2D_MA_NH} except that the photon index of the illuminating spectrum is assumed to be $\Gamma=1.0$, which is a typical value for an accreting neutron star \citep[See][]{Watanabe:2003} and is consistent with a value obtained by \citet{Suchy:2012}.
Figure~\ref{fig:EWk_fCS_GX301-2} shows how this constrains the two parameters, giving $N_\mathrm{H}=7.0 \pm 2.3$ and $A_\mathrm{metal}=0.66 ^{+0.41}_{-0.19}$.
A caveat of this analysis is that $f_\mathrm{CS}$ tends to be underestimated since the high energy region of the CS is difficult to measure due to overlap with the iron.
By a simple spectral simulation using the HETG detector response, $f_\mathrm{CS}$ can be underestimated by 80\%.

A more serious approach uses a numerical spectral model built by the Monte-Carlo simulations for data analysis.
This method is implemented with a table model in \texttt{XSpec} (we use version 12.9.0), and we can take account of an accurate detector response, which was not considered in the spectrum characterisation approach described above.
We fit the model which has free parameters of $N_\mathrm{H}$ and $A_\mathrm{metal}$ to the measured spectrum, fixing the photon index to $\Gamma=1.0$.
The fitted model is superposed in Figure~\ref{fig:GX301-2_spectrum} and it yields $N_\mathrm{H} = (1.078_{-0.128}^{+0.066}) \times 10^{24}\;\mathrm{cm}^{-2}$ and $A_\mathrm{metal} = 0.476_{-0.024}^{+0.049}$.
Structure of $\chi^2$ of the fitting and error contours for several confidence levels are shown in Figure~\ref{fig:contour}.

In summary, we constrained the two important parameters of the dense stellar wind in the GX 301$-$2 system, the column density $N_\mathrm{H}$ and the metal abundance $A_\mathrm{metal}$, with their small errors by using very limited energy range around the iron K$\alpha$ line and its CS.
The obtained values have good agreement with previous studies \citep{Watanabe:2003, Suchy:2012}.
The numerical model by the Monte-Carlo simulations provides us with a proper treatment of the data analysis, though the characterisation approach using $\mathrm{EW_{K\alpha}}$ and $f_\mathrm{CS}$ is also useful for quick estimation of the parameters.
The values obtained by the characterisation methods slightly differ from the fitting results.
This is mainly due to the underestimation of $f_\mathrm{CS}$, and therefore high energy resolution is required for the analysis.
In these methods, we needed to assume the spectral slope of the illuminating spectrum, which should be determined by a broadband spectrum.

\section{Conclusions}
\label{sec:conclusions}

We studied the basic nature of the CS by systematic evaluation of its dependence on spatial and temporal parameters.
The calculations are performed by Monte-Carlo simulations for sphere and slab geometries.
The dependence is obtained in a two-dimensional space of column density and metal abundance, demonstrating that the CS solves parameter degeneration between them which was seen in conventional spectral analysis using photoelectric absorption and fluorescence lines.
Unlike the iron fluorescence line, the CS is independent of spectral hardness of the illuminating spectrum.
The CS profile is highly dependent on the inclination angle of the slab geometry unless the slab is Compton-thick, and the time evolution of the CS is shown to be useful to constrain temporal information on the source if the intrinsic radiation is time variable.
Atomic binding of an electron in cold matter alters the scattering process, blurring the CS profile through the Doppler effect.
In practice, this effect resembles thermal broadening in a plasma with a moderate temperature of $\sim$5 eV.
Spectral diagnostics using the CS is available with high-resolution spectra obtained by grating spectrometers and microcalorimeters.

\section*{Acknowledgements}

This work is supported by JSPS KAKENHI grant number 24740190 and 24105007.
H.\;Y.\; is supported by the Advanced Leading graduate school for Photon Science (ALPS).
ACF acknowledges support from ERC Advanced Grant Feedback, 340442.

\appendix

\section{Semi-analytical Calculation of CS Profiles}
\label{sec:analytical_profile}

\begin{figure}
\begin{center}
\includegraphics[width=7.5cm]{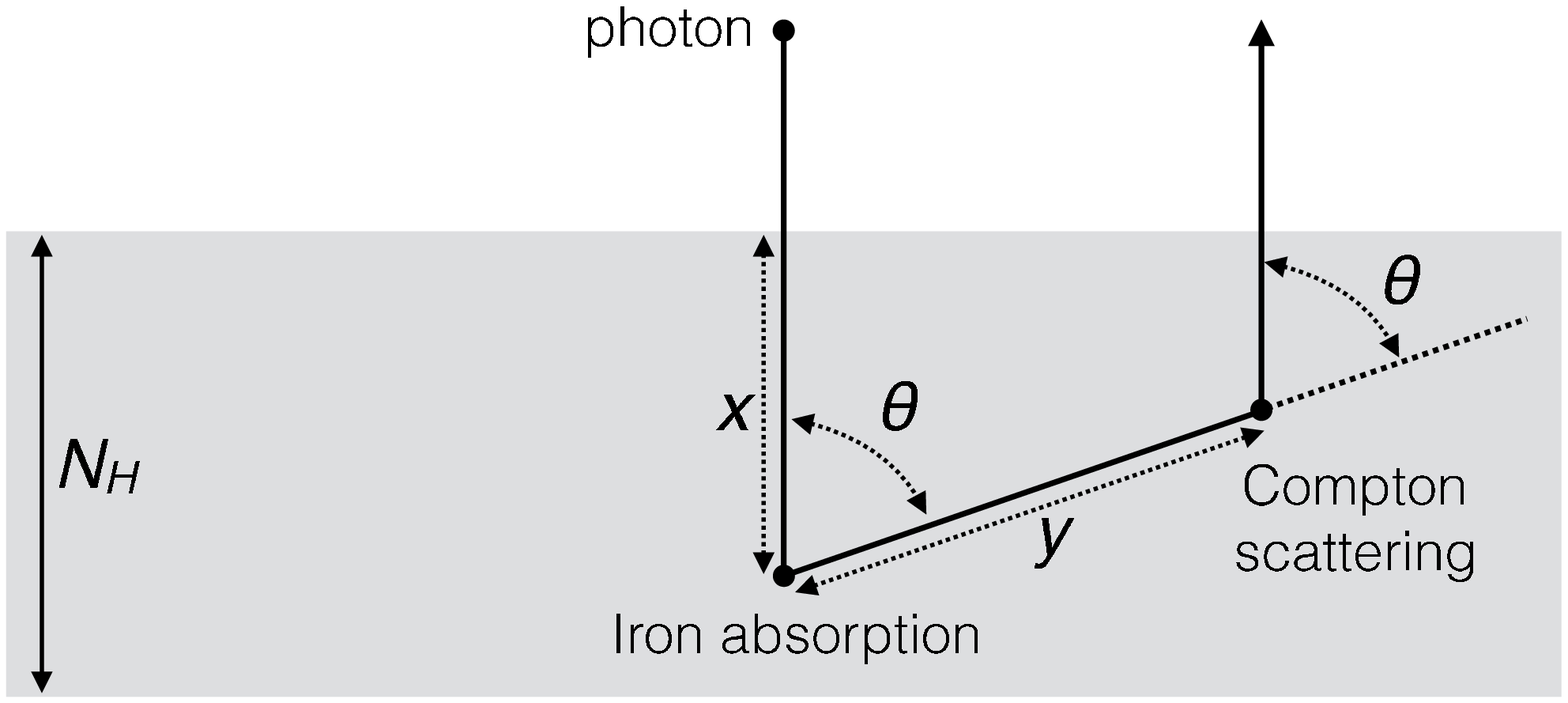}
\caption{Cross section view of the slab geometry in the semi-analytical calculation. We consider a photon into vertically downward which is absorbed in the slab, and then a fluorescence photon is emitted, resulting in Compton scattering to the vertically upward direction. $x$ is the depth at which the initial incident photon is absorbed by an iron atom. $y$ measures the path of the fluorescence photon following the absorption until it is Compton-scattered. $\theta$ denotes the scattering angle.}
\label{fig:calculation_geometry}
\end{center}
\end{figure}

\begin{figure}
\begin{center}
\includegraphics[width=7.5cm]{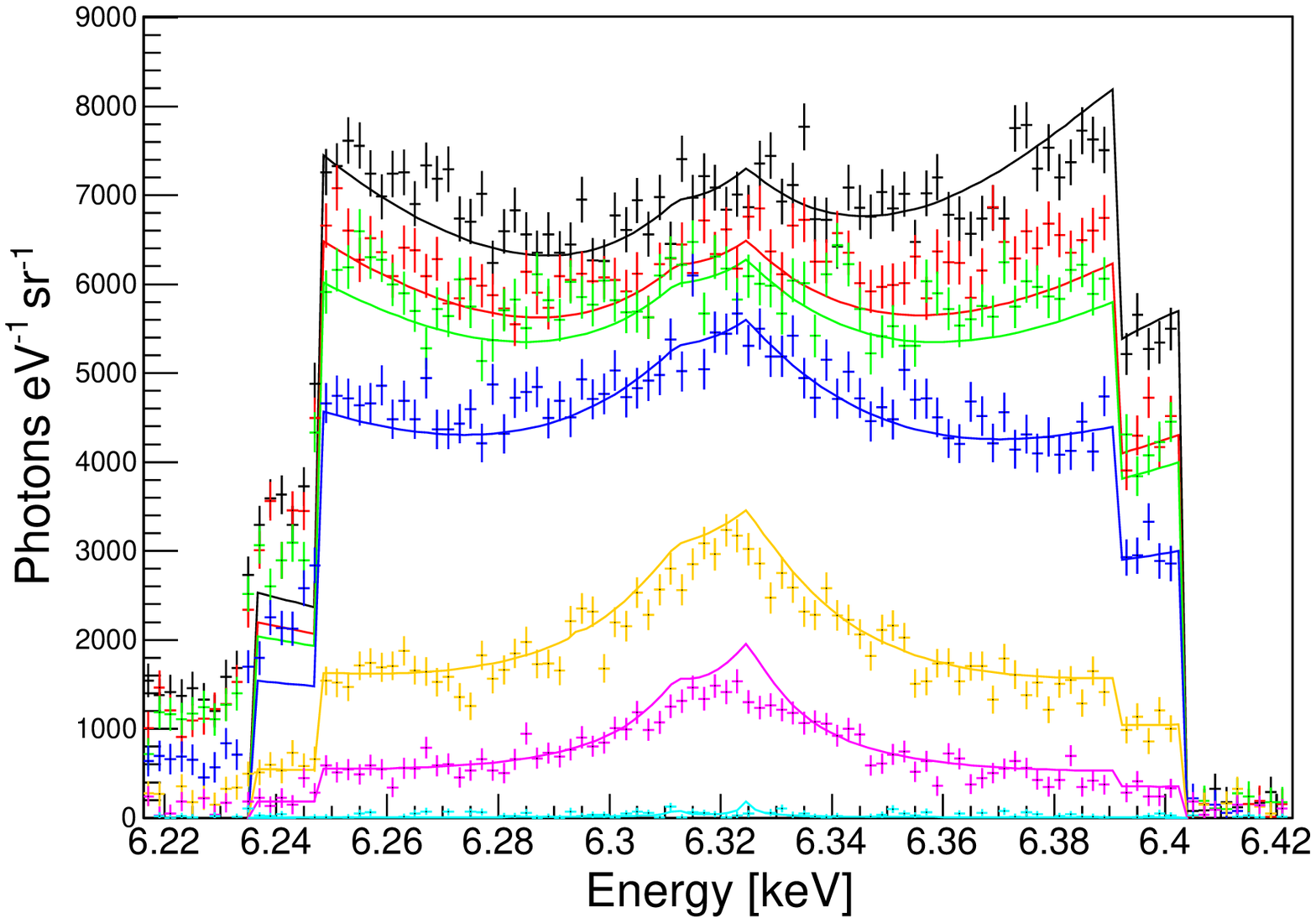}
\caption{CS profiles calculated by semi-analytical methods (solid lines) and by the Monte-Carlo simulations (data points). We assume a slab geometry, and the spectra are viewed from $\theta = 0^\circ$. We calculate them for several column densities: $N_\mathrm{H}=1\times 10^{22}\;\mathrm{cm^{-2}}$ (cyan), $1\times 10^{23}\;\mathrm{cm^{-2}}$ (magenta), $2\times 10^{22}\;\mathrm{cm^{-2}}$ (yellow), $5\times 10^{23}\;\mathrm{cm^{-2}}$ (blue), and $8\times 10^{23}\;\mathrm{cm^{-2}}$ (green), $1\times 10^{24}\;\mathrm{cm^{-2}}$ (red), $1\times 10^{25}\;\mathrm{cm^{-2}}$ (black), fixing a metal abundance at $A_\mathrm{metal} = 1.0$.}
\label{fig:calculation_comparison}
\end{center}
\end{figure}

We describe a semi-analytical calculation of the CS profile where we assume a slab geometry viewed from the vertically upward direction, namely $\theta = 0^\circ$.
This calculation can be directly compared with the Monte-Carlo results shown in Section~\ref{subsec:slab_angular}.
To calculate the CS profile, we primarily need to consider a photon which (1) goes into the slab vertically downward, (2) is absorbed and reprocessed into a fluorescence photon inside the slab, (3) is scattered to the vertically upward direction, and finally (4) escapes from the slab.
Such an event is drawn in Figure~\ref{fig:calculation_geometry}.
We define three probability distribution functions;
$P_1(E_0,x)$: a probability that the incident photon with an energy of $E_0$ is absorbed by an iron atom at a depth of $x$;
$P_2(x,y,\theta)$: a probability that an Fe K$\alpha$ photon emitted at a depth of $x$ is Compton-scattered after it runs for a length of $y$;
$P_3(x,y,\theta)$ is a probability that a scattered photon escapes from the slab without any interaction.
The CS profile (as a function of scattering angle $\theta$) is obtained by integrating
\begin{equation}\label{eq:semi-analytical}
P(\theta;E_0, x, y)=(1+\cos^2\theta) P_1(E_0, x) P_2(x,y,\theta) P_3(x,y,\theta)
\end{equation}
over $E_0$ in the incident spectrum and over $x$ and $y$ for allowed ranges in the slab geometry.
The factor of $(1+\cos^2\theta)$ comes from the Thomson differential cross section (Equation~\ref{eq:thomson}).

We performed integrations described above for several values of the column density.
In the calculation of $P_1(E_0, x)$, in addition to the direct path reaching the absorption point $x$ without any interaction, we included a photon path that has experienced Compton scattering once or twice before the absorption, and assumed that the path is restricted along the one-dimensional region (scattering angle is either 0$^\circ$ or 180$^\circ$).
This approximation is described in \citet{Basko:1978}.
In the calculation of $P_1(E_0, x)$, we also neglected the change of the photon energy via scatterings.
The calculation results given by (\ref{eq:semi-analytical}) are superposed on the Monte-Carlo results in Figure~\ref{fig:calculation_comparison}.
The semi-analytical solutions well agree with the Monte-Carlo calculations for the profile of the first-order CS, while discrepancy arises from second-order CS (scattered twice) at a low energy region, which is clearly seen particularly for high $N_\mathrm{H}$.

\bibliographystyle{mnras}
\bibliography{compton_shoulder}

\begin{thebibliography}{}
\makeatletter
\relax
\def\mn@urlcharsother{\let\do\@makeother \do\$\do\&\do\#\do\^\do\_\do\%\do\~}
\def\mn@doi{\begingroup\mn@urlcharsother \@ifnextchar [ {\mn@doi@}
  {\mn@doi@[]}}
\def\mn@doi@[#1]#2{\def\@tempa{#1}\ifx\@tempa\@empty \href
  {http://dx.doi.org/#2} {doi:#2}\else \href {http://dx.doi.org/#2} {#1}\fi
  \endgroup}
\def\mn@eprint#1#2{\mn@eprint@#1:#2::\@nil}
\def\mn@eprint@arXiv#1{\href {http://arxiv.org/abs/#1} {{\tt arXiv:#1}}}
\def\mn@eprint@dblp#1{\href {http://dblp.uni-trier.de/rec/bibtex/#1.xml}
  {dblp:#1}}
\def\mn@eprint@#1:#2:#3:#4\@nil{\def\@tempa {#1}\def\@tempb {#2}\def\@tempc
  {#3}\ifx \@tempc \@empty \let \@tempc \@tempb \let \@tempb \@tempa \fi \ifx
  \@tempb \@empty \def\@tempb {arXiv}\fi \@ifundefined
  {mn@eprint@\@tempb}{\@tempb:\@tempc}{\expandafter \expandafter \csname
  mn@eprint@\@tempb\endcsname \expandafter{\@tempc}}}

\bibitem[\protect\citeauthoryear{Agostinelli et~al.,}{Agostinelli
  et~al.}{2003}]{Agostinelli:2003}
Agostinelli S.,  et~al., 2003, Nuclear Instruments and Methods in Physics
  Research Section A: Accelerators, Spectrometers, Detectors and Associated
  Equipment, 506, 250

\bibitem[\protect\citeauthoryear{Allison et~al.,}{Allison
  et~al.}{2006}]{Allison:2006}
Allison J.,  et~al., 2006, IEEE Transactions on Nuclear Science, 53, 270

\bibitem[\protect\citeauthoryear{Anders \& Grevesse}{Anders \&
  Grevesse}{1989}]{Anders:1989}
Anders E.,  Grevesse N.,  1989, Geochim. Cosmochim. Acta, 53, 197

\bibitem[\protect\citeauthoryear{Barcons, Nandra, Barret, den Herder, Fabian,
  Piro, Watson  \& the Athena~team}{Barcons et~al.}{2015}]{Barcons:2015}
Barcons X.,  Nandra K.,  Barret D.,  den Herder J.-W.,  Fabian A.~C.,  Piro L.,
   Watson M.~G.,   the Athena~team 2015, Journal of Physics: Conference Series,
  610, 012008

\bibitem[\protect\citeauthoryear{Basko}{Basko}{1978}]{Basko:1978}
Basko M.,  1978, ApJ, 223, 268

\bibitem[\protect\citeauthoryear{Ert{\u u}gral, Apayd{\i}n, {\c C}evik, Ert{\u
  u}grul  \& Kobya}{Ert{\u u}gral et~al.}{2007}]{Ertugral:2007}
Ert{\u u}gral B.,  Apayd{\i}n G.,  {\c C}evik U.,  Ert{\u u}grul M.,   Kobya
  A.~{\. I}.,  2007, Radiation Physics and Chemistry, 76, 15

\bibitem[\protect\citeauthoryear{Furui, Fukazawa, Odaka, Kawaguchi, Ohno  \&
  Hayashi}{Furui et~al.}{2016}]{Furui:2016}
Furui S.,  Fukazawa Y.,  Odaka H.,  Kawaguchi T.,  Ohno M.,   Hayashi K.,
  2016, ApJ, 818, 164

\bibitem[\protect\citeauthoryear{Garc{\'\i}a, Kallman  \&
  Mushotzky}{Garc{\'\i}a et~al.}{2011}]{Garcia:2011}
Garc{\'\i}a J.,  Kallman T.~R.,   Mushotzky R.~F.,  2011, Astrophysical
  Journal, 731, 131

\bibitem[\protect\citeauthoryear{George \& Fabian}{George \&
  Fabian}{1991}]{George:1991}
George I.~M.,  Fabian A.~C.,  1991, Monthly Notices of the Royal Astronomical
  Society (ISSN 0035-8711), 249, 352

\bibitem[\protect\citeauthoryear{Hagino, Odaka, Done, Gandhi, Watanabe, Sako
  \& Takahashi}{Hagino et~al.}{2015}]{Hagino:2015}
Hagino K.,  Odaka H.,  Done C.,  Gandhi P.,  Watanabe S.,  Sako M.,   Takahashi
  T.,  2015, Monthly Notices of the Royal Astronomical Society, 446, 663

\bibitem[\protect\citeauthoryear{{Hitomi collaboration}}{{Hitomi
  collaboration}}{2016}]{Hitomi:2016}
{Hitomi collaboration} 2016, Nature, 535, 117

\bibitem[\protect\citeauthoryear{Ikeda, Awaki  \& Terashima}{Ikeda
  et~al.}{2009}]{Ikeda:2009}
Ikeda S.,  Awaki H.,   Terashima Y.,  2009, The Astrophysical Journal, 692, 608

\bibitem[\protect\citeauthoryear{Koyama, Maeda, Sonobe, Takeshima, Tanaka  \&
  Yamauchi}{Koyama et~al.}{1996}]{Koyama:1996}
Koyama K.,  Maeda Y.,  Sonobe T.,  Takeshima T.,  Tanaka Y.,   Yamauchi S.,
  1996, Publ. of the Astronomical Society of Japan, 48, 249

\bibitem[\protect\citeauthoryear{Koyama, Inui, Matsumoto  \& Tsuru}{Koyama
  et~al.}{2008}]{Koyama:2008}
Koyama K.,  Inui T.,  Matsumoto H.,   Tsuru T.~G.,  2008, Publications of the
  Astronomical Society of Japan, 60, 201

\bibitem[\protect\citeauthoryear{Krause}{Krause}{1979}]{Krause:1979}
Krause M.~O.,  1979, Journal of Physical and Chemical Reference Data, 8, 307

\bibitem[\protect\citeauthoryear{Leahy \& Creighton}{Leahy \&
  Creighton}{1993}]{Leahy:1993}
Leahy D.~A.,  Creighton J.,  1993, R.A.S. MONTHLY NOTICES V.263, 263, 314

\bibitem[\protect\citeauthoryear{Matt}{Matt}{2002}]{Matt:2002}
Matt G.,  2002, Monthly Notice of the Royal Astronomical Society, 337, 147

\bibitem[\protect\citeauthoryear{Matt, Perola  \& Piro}{Matt
  et~al.}{1991}]{Matt:1991}
Matt G.,  Perola G.~C.,   Piro L.,  1991, Astronomy and Astrophysics (ISSN
  0004-6361), 247, 25

\bibitem[\protect\citeauthoryear{Mori et~al.,}{Mori et~al.}{2015}]{Mori:2015}
Mori K.,  et~al., 2015, The Astrophysical Journal, 814, 94

\bibitem[\protect\citeauthoryear{Murphy \& Yaqoob}{Murphy \&
  Yaqoob}{2009}]{Murphy:2009}
Murphy K.~D.,  Yaqoob T.,  2009, Monthly Notices of the Royal Astronomical
  Society, 397, 1549

\bibitem[\protect\citeauthoryear{Odaka, Aharonian, Watanabe, Tanaka, Khangulyan
   \& Takahashi}{Odaka et~al.}{2011}]{Odaka:2011}
Odaka H.,  Aharonian F.,  Watanabe S.,  Tanaka Y.,  Khangulyan D.,   Takahashi
  T.,  2011, The Astrophysical Journal, 740, 103

\bibitem[\protect\citeauthoryear{Odaka, Khangulyan, Tanaka, Watanabe, Takahashi
   \& Makishima}{Odaka et~al.}{2014}]{Odaka:2014}
Odaka H.,  Khangulyan D.,  Tanaka Y.~T.,  Watanabe S.,  Takahashi T.,
  Makishima K.,  2014, The Astrophysical Journal, 780, 38

\bibitem[\protect\citeauthoryear{Revnivtsev et~al.,}{Revnivtsev
  et~al.}{2004}]{Revnivtsev:2004}
Revnivtsev M.~G.,  et~al., 2004, Astronomy and Astrophysics, 425, L49

\bibitem[\protect\citeauthoryear{Reynolds et~al.,}{Reynolds
  et~al.}{2014}]{Reynolds:2014}
Reynolds C.,  et~al., 2014, arXiv

\bibitem[\protect\citeauthoryear{Ross \& Fabian}{Ross \&
  Fabian}{2005}]{Ross:2005}
Ross R.~R.,  Fabian A.~C.,  2005, Monthly Notices of the Royal Astronomical
  Society, 358, 211

\bibitem[\protect\citeauthoryear{Smith, Odaka, Audard  \& Brown}{Smith
  et~al.}{2014}]{Smith:2014}
Smith R.~K.,  Odaka H.,  Audard M.,   Brown G.~V.,  2014, arXiv

\bibitem[\protect\citeauthoryear{Suchy, F{\"u}rst, Pottschmidt, Caballero,
  Kreykenbohm, Wilms, Markowitz  \& Rothschild}{Suchy
  et~al.}{2012}]{Suchy:2012}
Suchy S.,  F{\"u}rst F.,  Pottschmidt K.,  Caballero I.,  Kreykenbohm I.,
  Wilms J.,  Markowitz A.,   Rothschild R.~E.,  2012, Astrophysical Journal,
  745, 124

\bibitem[\protect\citeauthoryear{Sunyaev \& Churazov}{Sunyaev \&
  Churazov}{1996}]{Sunyaev:1996}
Sunyaev R.~A.,  Churazov E.~M.,  1996, Astronomy Letters, 22, 648

\bibitem[\protect\citeauthoryear{Sunyaev \& Churazov}{Sunyaev \&
  Churazov}{1998}]{Sunyaev:1998}
Sunyaev R.,  Churazov E.,  1998, Monthly Notices of the Royal Astronomical
  Society, 297, 1279

\bibitem[\protect\citeauthoryear{Sunyaev, Markevitch  \& Pavlinsky}{Sunyaev
  et~al.}{1993}]{Sunyaev:1993}
Sunyaev R.~A.,  Markevitch M.,   Pavlinsky M.,  1993, Astrophysical Journal,
  407, 606

\bibitem[\protect\citeauthoryear{Sunyaev, Uskov  \& Churazov}{Sunyaev
  et~al.}{1999}]{Sunyaev:1999}
Sunyaev R.~A.,  Uskov D.~B.,   Churazov E.~M.,  1999, Astronomy Letters, 25,
  199

\bibitem[\protect\citeauthoryear{Takahashi et~al.,}{Takahashi
  et~al.}{2014}]{Takahashi:2014}
Takahashi T.,  et~al., 2014, SPIE Astronomical Telescopes + Instrumentation,
  9144, 914425

\bibitem[\protect\citeauthoryear{Thompson et~al.}{Thompson
  et~al.}{2001}]{Thompson:2001}
Thompson A.~C.,  et~al., 2001, X-ray Data Booklet (Berkeley, CA: Lawrence
  Berkeley National Laboratory)

\bibitem[\protect\citeauthoryear{Vainshtein, Sunyaev  \& Churazov}{Vainshtein
  et~al.}{1998}]{Vainshtein:1998}
Vainshtein L.~A.,  Sunyaev R.,   Churazov E.~M.,  1998, Astronomy Letters, 24,
  271

\bibitem[\protect\citeauthoryear{Watanabe et~al.,}{Watanabe
  et~al.}{2003}]{Watanabe:2003}
Watanabe S.,  et~al., 2003, The Astrophysical Journal, 597, L37

\bibitem[\protect\citeauthoryear{Watanabe et~al.,}{Watanabe
  et~al.}{2006}]{Watanabe:2006}
Watanabe S.,  et~al., 2006, The Astrophysical Journal, 651, 421

\bibitem[\protect\citeauthoryear{Yaqoob}{Yaqoob}{2012}]{Yaqoob:2012}
Yaqoob T.,  2012, Monthly Notices of the Royal Astronomical Society, 423, 3360

\bibitem[\protect\citeauthoryear{Yaqoob \& Murphy}{Yaqoob \&
  Murphy}{2010}]{Yaqoob:2010}
Yaqoob T.,  Murphy K.~D.,  2010, Monthly Notices of the Royal Astronomical
  Society, 412, 277

\bibitem[\protect\citeauthoryear{Zhang et~al.,}{Zhang
  et~al.}{2015}]{Zhang:2015}
Zhang S.,  et~al., 2015, The Astrophysical Journal, 815, 132

\makeatother
\end{thebibliography}


\end{document}